\documentclass[prb,twocolumn,amsmath,amssymb,floatfix,superscriptaddress,aps]{revtex4-2}

\usepackage{braket}
\usepackage{tikz}
 \usepackage[makeroom]{cancel}

\usepackage{hyperref}
\usepackage[capitalise]{cleveref}
\usepackage[normalem]{ulem}
\usepackage{geometry}
\newcommand{\etan}{\ensuremath{\eta\sigma}}
\newcommand{\etab}{\ensuremath{\bar{\eta}\,\bar{\sigma}}}

\usetikzlibrary{arrows.meta}   
\tikzset{> = {Stealth[length=7pt,width=5pt]}}

\newcommand{\RTDone}[6]{%
	\mathrel{
		\tikz[baseline=1.5ex,scale=0.8]{%
			\draw[-] (0,1) -- (2,1)
			node[pos=0,left]  {$\scriptstyle #1$}
			node[pos=1,right] {$\scriptstyle #2$};
			
			\draw[-] (0,0) -- (2,0)
			node[pos=0,left]  {$\scriptstyle #4$}
			node[pos=1,right] {$\scriptstyle #5$};
			
			\draw[#6] (0,1) -- (2,0)
			node[midway,right] {$\scriptstyle #3$};
		}%
	}%
}

\newcommand{\RTDtwo}[6]{%
	\mathrel{
		\tikz[baseline=1.5ex,scale=0.8]{%
			\draw[-] (0,1) -- (2,1)
			node[pos=0,left]  {$\scriptstyle #1$}
			node[pos=1,right] {$\scriptstyle #2$};
			
			\draw[-] (0,0) -- (2,0)
			node[pos=0,left]  {$\scriptstyle #4$}
			node[pos=1,right] {$\scriptstyle #5$};
			
			\draw[#6] (0,0) -- (2,1)
			node[midway,left] {$\scriptstyle #3$};
		}%
	}%
}

\newcommand{\RTDthree}[7]{%
	\mathrel{%
		\tikz[baseline=1.5ex,scale=0.8]{%
			\draw[-] (0,1) -- (2,1)
			node[pos=0,left]  {$\scriptstyle #1$}
			node[midway,above] {$\scriptstyle #2$}
			node[pos=1,right] {$\scriptstyle #3$};
			
			\draw[-] (0,0) -- (2,0)
			node[pos=0,left]  {$\scriptstyle #5$}
			node[pos=1,right] {$\scriptstyle #6$};
			
			\draw[#7] (0,1)
			to[out=-90,in=-90]          
			node[midway,above] {$\scriptstyle #4$}
			(2,1);
		}%
	}%
}

\newcommand{\RTDfour}[7]{%
	\mathrel{%
		\tikz[baseline=1.5ex,scale=0.8]{%
			\draw[-] (0,1) -- (2,1)
			node[pos=0,left]  {$\scriptstyle #1$}
			node[pos=1,right] {$\scriptstyle #2$};
			
			\draw[-] (0,0) -- (2,0)
			node[pos=0,left]  {$\scriptstyle #4$}
				node[midway,below] {$\scriptstyle #5$}
			node[pos=1,right] {$\scriptstyle #6$};
			
			\draw[#7] (0,0)
			to[out=90,in=90]                                
			node[midway,below] {$\scriptstyle #3$} 
			(2,0);
		}%
	}%
}

\newcommand{\ud}{\uparrow\downarrow}
\newcommand{\du}{\downarrow\uparrow}
\newcommand{\uu}{\uparrow\uparrow}
\newcommand{\dd}{\downarrow\downarrow}

\begin{document}
\title{Current precision in  interacting hybrid Normal-Superconducting systems}
\author{Nahual Sobrino}\email{nsobrino@ictp.it}
\affiliation{The Abdus Salam International Center for Theoretical Physics, Strada Costiera 11, 34151 Trieste, Italy}
\author{Fabio Taddei}
\affiliation{NEST, Istituto Nanoscienze-CNR and Scuola Normale Superiore, Piazza San Silvestro 12, I-56127 Pisa, Italy}
\author{Rosario Fazio} 
\affiliation{The Abdus Salam International Center for Theoretical Physics, Strada Costiera 11, 34151 Trieste, Italy}
\affiliation{Dipartimento di Fisica, Universit\`a di Napoli ``Federico II”, Monte S. Angelo, I-80126 Napoli, Italy}
\author{Michele Governale}
\affiliation{School of Chemical and Physical Sciences and MacDiarmid Institute for Advanced Materials and Nanotechnology,
	Victoria University of Wellington, PO Box 600, Wellington 6140, New Zealand}

\begin{abstract}
	We study Andreev-mediated transport and current fluctuations in interacting normal--superconducting quantum-dot systems. Using a generalized master equation based on real-time diagrammatics and full counting statistics, we compute the steady-state current, zero-frequency noise, and rate of entropy production in the large superconducting-gap limit. 
	We show how Coulomb interactions modify Andreev-mediated transport by renormalizing resonant conditions and suppressing superconducting coherence, leading to a pronounced reduction of current precision even when average currents are only weakly affected. 
	These effects are particularly evident at high temperatures, where conventional Coulomb-blockade features are thermally smeared while fluctuation properties remain highly sensitive. 
	By analyzing thermodynamic uncertainty relations, we demonstrate that violations of the quantum bound present in the noninteracting regime are progressively reduced and eventually suppressed as interactions increase, whereas the recently proposed hybrid bound remains satisfied. 
	Our results clarify how
     Coulomb interactions, and nonequilibrium fluctuations jointly determine transport properties in hybrid superconducting devices, and establish current precision as a robust benchmark for interacting Andreev transport beyond the noninteracting limit.
\end{abstract}
 \date{\today}
 \maketitle
 
 \allowdisplaybreaks

The ability to generate electric currents with high precision is essential in several contexts. One notable example arises in metrology, where the current standard
can be defined, using quantum-based devices, producing relatively large currents with a relative uncertainty no greater than $10^{-8}$~\cite{Kaneko2016,Kaneko2024}.
 Another important example concerns the operation of heat engines, for which high thermodynamic efficiency is desirable, but only if accompanied by a sufficiently large and well-controlled power output~\cite{Benenti2017,Cangemi2024,Balduque2025,sobrino2025thermoelectric}. This requirement becomes especially stringent in nanoscale devices, where the achievable currents are inherently small. The Thermodynamic Uncertainty Relation (TUR) establishes a fundamental bound linking the precision of nonequilibrium currents to entropy production~\cite{Pietzonka2018},
thereby limiting how accurately currents can be controlled in small-scale systems.
Originally derived for classical Markovian dynamics~\cite{barato2015thermodynamic}, these relations formalize the intuitive tradeoff between stability and dissipation: reducing current fluctuations necessarily comes at the cost of increased entropy production. Over the past decade, TURs have emerged as a powerful framework to characterize nonequilibrium processes across a wide range of physical systems, from biomolecular machines to mesoscopic electronic conductors~\cite{ Ehrlich2021,  Guarnieri2019, Kheradsoud2019, Proesmans2019, Pietzonka2018, Prech2023, Ptaszynski2018, Saryal2019, Saryal2021, Brandner2018, Saryal2022, gingrich2016dissipation, taddei2023thermodynamic, horowitz2020thermodynamic, Kamijima2021, Lopez2023, Misaki2021, Palmqvist2024, Wozny2025,Zhang2025, Potanina2021,Lu2022}.
Recently,
a TUR that explicitly accounts for coherent transport has been 
 derived within the framework of scattering theory. This quantum TUR hold for non-interacting, phase-coherent conductors~\cite{Brandner2025}.

Hybrid normal--superconducting nanostructures provide a particularly rich platform to explore the possibility to enhance current precision for a given rate of entropy production.
In such systems, transport in the subgap regime is governed by Andreev reflection processes, which intrinsically couple electron and hole degrees of freedom and are underpinned by macroscopic superconducting coherence~\cite{Degennesbook,Blonder1982,braggio2011superconducting,eldridge2010superconducting}.
As a result, hybrid devices can exhibit fluctuation properties that differ qualitatively from those of purely normal conductors, leading to various interesting phenomena.
 While the dissipationless supercurrent in a bulk superconductor is a ground-state property, and
therefore is not accompanied by any fluctuations (noiseless), the introduction of a normal-metal interface changes this situation.
In such hybrid structures, current fluctuations arise from the interplay between the single-particle statistics of the normal metal and the collective macroscopic coherence of the superconductor, as first discussed in Refs.~\cite{Khlus1987,deJong1994,Muzykantskii1994,Martin1996,anantram1996current}.
In particular, these fluctuations serve as a powerful diagnostic tool to probe topological bound states (see Refs.~\cite{Bolech2007,Golub2011,WU2012,Lu2012} for early works),
and the underlying symmetry of the superconducting order parameter~\cite{Weiss2017,Seoane2020,Arrachea2024,Gonzalez2025}.


Several recent works have highlighted the role of Andreev processes in enhancing current precision, by analyzing the classical~\cite{Misaki2021,Manzano2023Oct,Lopez2023,taddei2023thermodynamic,Ohnmacht2024} and the quantum~\cite{mayo2025thermodynamic,ohnmacht2025role} TUR in superconducting junctions.
Violations of TURs were traced back to macroscopic superconducting coherence and the resulting Andreev-dominated transport, establishing a direct link between superconducting correlations and enhanced current precision even in regimes of significant thermal broadening.
Moreover, in a recent study~\cite{mayo2025thermodynamic}, a version of the quantum TUR for non-interacting hybrid superconducting system in the limit of an infinite superconducting gap was proposed.

 An important open question is how robust these precision-enhancing effects are once electron--electron interactions are taken into account. In realistic quantum-dot devices, Coulomb interactions are often comparable to or larger than tunnel couplings and can strongly modify both transport and fluctuation properties. While interactions are known to suppress superconducting proximity effects and renormalize Andreev resonances~\cite{Fazio1998,Clerk2000,governale2008real,pala2007nonequilibrium}, their impact on current precision and on the validity of the TURs in hybrid systems remains largely unexplored. Addressing this issue requires theoretical approaches capable of treating interactions beyond mean-field. 
 
 In this work, we address this problem by studying an interacting central region coupled to normal and superconducting electronic reservoirs using real-time diagrammatics combined with full counting statistics. This framework allows us to compute the steady-state current, zero-frequency noise, and rate of entropy production in the large-gap limit, treating the Coulomb interaction exactly within a reduced density-matrix formalism. We analyze the case in which the central region corresponds to a single quantum dot and a  double quantum dot, thereby capturing the combined effects of local and nonlocal Andreev processes.  We investigate systematically how interactions modify transport resonances, suppress superconducting coherence, and affect current precision.
 
The paper is organized as follows. 
In Section~\ref{Sectiontwo} we introduce the general transport setup and present the generalized master equation formalism used to compute the steady-state current and zero-frequency noise. 
Sections~\ref{Sectionthree} and~\ref{Sectionfour} are devoted to the single quantum dot and Cooper-pair–splitter (CPS) systems, respectively, where we analyze the current, noise, and violations of the quantum TUR for representative parameter regimes. 
Our conclusions are presented in Section~\ref{Sectionfive}. 
Appendix~\ref{Appendixone} provides the detailed derivation of the rate matrix within the real-time diagrammatic approach. 
In Appendix~\ref{Appendixtwo} we derive the Green’s functions within the equation-of-motion (EOM) method at the Hartree–Fock level for both the single quantum dot and the CPS, and we include a comparison between the generalized master equation and Green’s function results in the limits of small bias and weak interactions.

 \section{Model and Formalism}
 \label{Sectiontwo}
\begin{figure}
	\centering
	\includegraphics[width=\linewidth]{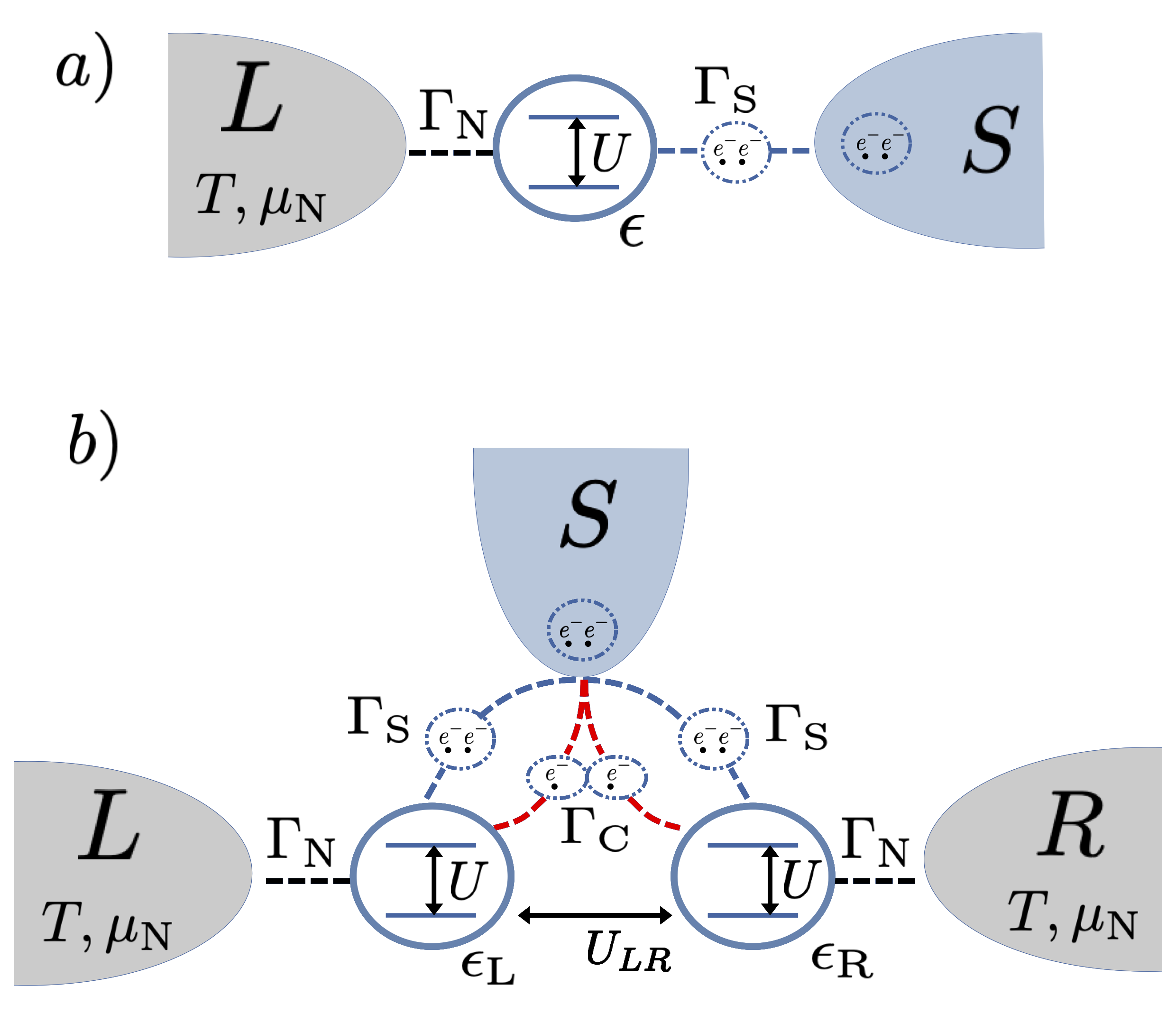}
	\caption{Schematic setup representation of the two interacting hybrid normal-superconducting systems studied: (a) Single quantum dot, and (b) Cooper-pair splitter. The blue dashed line corresponds to the local tunneling rate $\Gamma_S$, and the red dashed line corresponds to the nonlocal tunneling rate $\Gamma_C$.}
	\label{figure1}
\end{figure}

 We consider an interacting central region attached to a superconducting electrode and one or two normal electrodes, see Fig.~\ref{figure1}. The superconducting electrode couples to the central region through a local tunneling rate $\Gamma_S$ (blue dashed line), transferring two electrons locally, and through a nonlocal tunneling rate $\Gamma_C$ (red dashed line), which splits the two electrons between different sites.
 
 The superconducting electrode is grounded, and we assume that the superconducting gap is the largest energy scale, so we consider the limit $\Delta\to\infty$. In this limit, the quasi-particles in the superconductor do not contribute and the transport processes  between the superconductor and the central region can be included as an effective pairing term in the Hamiltonian of the central region. The form of the effective Hamiltonian will be discussed when  the two different systems are studied.  The leads are considered to be in local thermal equilibrium, with Hamiltonian 
$H_{\rm leads}=\sum_{\eta k \sigma}\epsilon_{\eta  k i} c^{\dagger}_{\eta k \sigma} c_{\eta  k \sigma}$ for $\eta={\rm L,R}$ and with $ c^{\dagger}_{\eta k \sigma}\quad ( c_{\eta k \sigma})$ the creation (annihilation) operator for an electron with spin $\sigma$ in state $k$ of lead $\eta$.
The leads are coupled to the central region by means of the tunneling Hamiltonians  
$H_{{\rm tunn}} =\sum_{\eta,i}\sum_{k,\sigma}  V_{\eta,i}\left(c^\dagger_{\eta k\sigma}d_{i,\sigma} + \text{H.c.}\right)$, with $V_{\eta i}$ the tunnel-coupling amplitude between the dot $i$ of the central region and the normal lead $\eta$. For the sake of simplicity, in the case when there are two normal leads, we assume $V_{L,L}=V_{R,R}=V_N$ and $V_{L,R}=V_{R,L}=0$. The symbol  $ d^{\dagger}_{i \sigma}\quad ( d_{i\sigma})$ denotes the creation (annihilation) operator for an electron with spin $\sigma$ in dot $i$. We adopt the wide-band approximation for which the density of states of the normal leads at the Fermi energy $\rho_N$ is constant in the relevant energy range, so that the tunneling coupling strength $\Gamma_N = 2\pi \rho_N |V_N|^2$ is energy independent.

The dynamics of the central region is determined by its reduced density matrix $\rho_{c}$, with elements 
$P^{\xi_i}_{\xi_j}\equiv \bra{\xi_j}\rho_c\ket{\xi_i}$,
being $\ket{\xi_\alpha}$ an element of the many-body basis $\{\ket{\xi}\}$ for the isolated central region. The time evolution of $\rho_{c}$ is governed by the generalized  master equation ($\hbar=1$)
 \begin{align*}
	\frac{d}{dt}P^{\xi_1}_{\xi_2}(t)+i\delta E_{\xi_1,\xi_2}P^{\xi_1}_{\xi_2}(t)=\sum_{\xi_{1}',\xi_{2}'}\int_{t_0}^{t}dt'W^{\xi_1\xi_1'}_{\xi_2\xi_2'}(t,t')P^{\xi_1'}_{\xi_2'}(t'),
\end{align*}
\normalsize
where $\delta E_{\xi_1,\xi_2} = E_{\xi_1}-E_{\xi_2}$. In the steady-state regime, $\rho_c$ is independent of $t$ and the equation can be rewritten as
 \begin{align}
	i\delta E_{\xi_1,\xi_2}P^{\xi_1}_{\xi_2}=\sum_{\xi_{1}',\xi_{2}'}	W^{\xi_1\xi_1'}_{\xi_2\xi_2'}P^{\xi_1'}_{\xi_2'},
	\label{ME}
\end{align}
where  $W^{\xi_1\xi_1'}_{\xi_2\xi_2'}\equiv \int_{-\infty}^{t}dt'W^{\xi_1\xi_1'}_{\xi_2\xi_2'}(t,t')$ are the generalized transition rates. The normalization condition implies $\sum_{\xi} P^{\xi}_{\xi}=1$. 
 In order to compute the transition rates we employ a real time diagrammatics technique \cite{governale2008real,konig1996zero}, which allows for a systematic expansion in the coupling Hamiltonian while keeping the exact interaction dependence. In the following sections we compute the transition rates  up to linear order in the coupling strengths to the normal lead $\Gamma_N$, and the local and non-local tunneling rates $\Gamma_S$ and $\Gamma_C$.
 

 To determine the stationary current cumulants we consider the full counting statistics. We introduce a counting field $\chi$ for the normal leads and dress the corresponding jump terms: a transition that transfers one electron into (out of) that lead acquires a phase factor $e^{+i\chi}$ ($e^{-i\chi}$). This defines the $\chi$–dependent rate kernel $\mathcal{W}(\chi)$.
\begin{figure*}
	\centering
	\includegraphics[width=\linewidth]{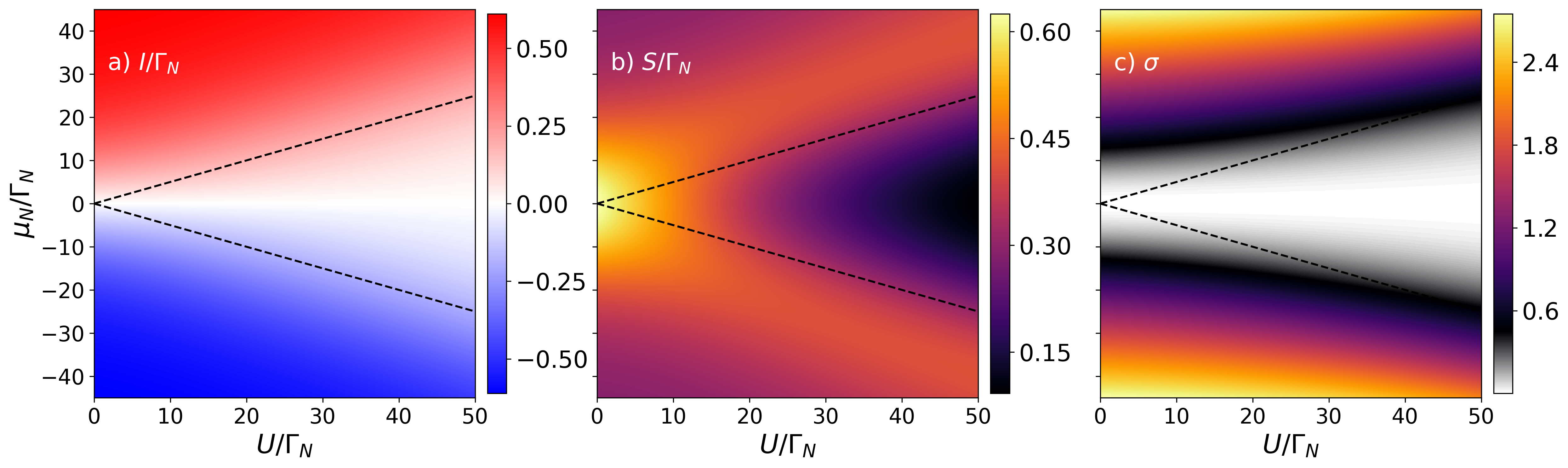}
	\caption{(a) Superconducting current, (b) Noise, and (c) Rate of entropy production in the single dot setup as a function of the Coulomb interaction $U$ and the chemical potential of the normal lead $\mu_N$.  The dashed black line represents the Coulomb blockade threshold occurring at $\mu_N=-U/2$. The parameters are $\varepsilon=-U/2$, $k_B T=10\Gamma_N$ and $\Gamma_S=\sqrt{5/3}\Gamma_N$. Energies are in units of $\Gamma_N$.}
	\label{figure2}
\end{figure*}
Throughout this section we adopt the superoperator (superspace) notation introduced in Ref.~\cite{flindt2010counting}, in which $\mathcal{W}(\chi)$ acts \emph{linearly} on the reduced density operator. Operators are vectorized as $|X\rangle\!\rangle \equiv \hat X$, where $\hat X$ is a conventional quantum operator, and $|X\rangle\!\rangle$ is the corresponding ket in the superspace. The inner product between bras and kets is defined as $	\langle\!\langle A|B\rangle\!\rangle \equiv \mathrm{Tr}\!\left(\hat{A}^\dagger \hat{B}\right)$.

 The reduced density operator is represented as $|0\rangle\!\rangle \equiv \hat \rho_{c}$ and the identity operator $\mathbb{\hat I}$ in the conventional Hilbert space as $\langle\!\langle \mathbb{I}|$, hence for $\chi=0$:

\begin{align}
	\mathcal{W}(0)\,|0\rangle\!\rangle &= 0, &
	\langle\!\langle \mathbb{I}|\,\mathcal{W}(0) &= 0, &
	\langle\!\langle \mathbb{I}|0\rangle\!\rangle &= 1 .
\end{align}

Derivatives of the kernel at $\chi=0$ are denoted by
\begin{align}
	\mathcal{W}^{(n)} \;\equiv\;
	\left.\frac{\partial^{n}\mathcal{W}(\chi)}{\partial (i\chi)^{n}}\right|_{\chi=0},
\end{align}
which are straightforward to obtain once the generalized transition rates are known since the counting field enters only through the factors $e^{\pm i\chi}$. With the projectors $P=|0\rangle\!\rangle\langle\!\langle\mathbb{I}|$ and $Q=1-P$,  one can define the Drazin pseudoinverse restricted to the $Q$–subspace as 
$\mathcal{R} \;=\; Q\,\mathcal{W}(0)^{-1} Q$. 
The steady-state current $I$ and zero-frequency noise $S$ then read \cite{flindt2010counting}
\begin{subequations}
\begin{align}
	I &= \langle\!\langle \mathbb{I}| \mathcal{W}^{(1)}\,|0\rangle\!\rangle, \label{eq:I_mean}\\[3pt]
	S &= \langle\!\langle \mathbb{I}| \Big[\, \mathcal{W}^{(2)} - 2\,\mathcal{W}^{(1)} \mathcal{R}\, \mathcal{W}^{(1)} \,\Big] |0\rangle\!\rangle. \label{eq:S_zero}
\end{align}
\label{eq_I_S}
\end{subequations}
Since we consider the same counting field for both normal leads, the current cumulants computed through  Eq.~(\ref{eq_I_S}) correspond to the sum of the currents in the normal leads and hence, due to current conservation, to the current flowing in the superconducting lead.

Having access to the steady-state current, zero-frequency noise, and rate of entropy production, we can assess the precision of charge transport through the TURs, which provide universal bounds linking current fluctuations to dissipation. In the present context, TURs offer a compact way to quantify how superconducting coherence and Coulomb interactions affect transport precision in hybrid normal--superconducting systems. We therefore consider the classical, quantum, and hybrid quantum TURs, which are respectively relevant for incoherent Markovian dynamics, phase-coherent noninteracting conductors, and Andreev-dominated transport
\begin{subequations}
\begin{align}
\mathcal{F}&\equiv \frac{Fe\sigma}{2|I|}-1\ge 0,\qquad \qquad\qquad\text{classical}\\
\mathcal{Q}&\equiv F\sinh\left(\frac{e\sigma}{2|I|}\right)-1\ge 0,\qquad \text{quantum}\\
\mathcal{Q_H}&\equiv \frac{F}{2}\sinh\left(\frac{e\sigma}{|I|}\right)-1\ge 0.\qquad \text{hybrid quantum}
\end{align}
\label{TURs}
\end{subequations}
Here $F = S/(e|I|)$ is the Fano factor and $\sigma$ denotes the rate of entropy production, which in our case is given by $\sigma = VI/T$, where $V = -\mu_N/e$ is the voltage applied to the normal leads.  In a recent study~\cite{mayo2025thermodynamic}, it was shown using an exact nonequilibrium Green’s function (NEGF) approach, that in the noninteracting limit a central region coupled to normal and superconducting leads can violate both the classical and quantum TURs. 
 These violations were found to occur for comparable tunnel couplings to the normal and superconducting reservoirs, with the strongest departures from the bounds appearing around $\Gamma_S \sim \sqrt{5/3}\,\Gamma_N$, and at temperatures $k_B T$ of the order of or larger than the tunnel broadenings. Since the quantum TUR provides the relevant benchmark for coherent noninteracting conductors, in the following we mainly focus on violations of the quantum inequality in this parameter regime.

  \section{Single Quantum Dot}
  \label{Sectionthree}
  When the central region is a quantum dot with 
  one spin-degenerate localized level
  as in Fig.~\ref{figure1} (a), it is described by the  Hamiltonian of the Anderson model
  \begin{align}
  	H_{QD}= \varepsilon\sum_{\sigma}n_{\sigma}+Un_{\uparrow}n_{\downarrow} ,
  \end{align}
  where $n_{\sigma}=d^{\dagger}_\sigma d_{\sigma}$ is the spin-resolved  occupation-number operator, $\varepsilon$ is the energy level, and $U$ is the energy cost of double occupancy due to the  Coulomb repulsion between the electrons. The effective full Hamiltonian of the system including the central regions contacted  to one normal and one superconducting lead in the $\Delta\to\infty$ limit reads \cite{droste2012josephson}
  \begin{align}
H=H_{QD}+H_{\rm leads}+H_{\rm tunn}-\frac{\Gamma_S}{2}\left(d^\dagger_\uparrow d^\dagger_\downarrow+{\rm H.c.}\right) ,
\label{eq_QD}
  \end{align}
where the last term accounts for the local Andreev reflection, being $\Gamma_S$ the tunneling rate between the superconductor and the central dot.
  
   The basis of the Fock space associated to the single dot $\{\ket{0},\ket{\uparrow},\ket{\downarrow},\ket{D}\equiv d^\dagger_\uparrow d^\dagger_\downarrow\ket{0}\}$ determines the transition rates needed to compute the density matrix by means of Eq.~(\ref{ME}).
    We calculate the rates $W^{\xi_1\xi_1'}_{\xi_2\xi_2'}$ to first order in  $\Gamma_N$ and for vanishing detuning between the energy of the empty and doubly occupied states, i.e., $\delta =E_{0}-E_{D}=0$. 
     This approximation captures both the transfer of charges through the tunneling barriers as well as energy-renormalization terms and is valid as long as $\Gamma_S,\Gamma_N,\delta\ll k_B T$.  A similar derivation of the rates for the case of two superconducting leads can be found in Ref.~\cite{pala2007nonequilibrium}. In our case, the rates are given by
  \begin{subequations}
\begin{align}
	W_{\sigma,D} &= \Gamma_N\, f^{-}(\frac{U}{2})e^{i \chi},\quad 	W_{D,\sigma} = \Gamma_N\, f^{+}(\frac{U}{2})e^{-i \chi},\\
	W_{\sigma,0} &= \Gamma_N\, f^{+}(-\frac{U}{2})e^{-i \chi},\quad W_{0,\sigma} = \Gamma_N\, f^{-}(-\frac{U}{2})e^{i \chi},\\
	W_{0,0} &= -\,2\,W_{\sigma,0}\big|_{\chi=0},\quad W_{D,D} = -\,2\,W_{\sigma,D}|_{\chi=0},\\
	W_{\sigma,\sigma} &= -\big( W_{D,\sigma} + W_{0,\sigma} \big)|_{\chi=0},\\
	W^{D,D}_{0,0} &= -\,\Gamma_N\!\left[\varphi^{-}(\frac{U}{2},+1)
	+ \varphi^{+}(-\frac{U}{2},-1)\right],\\
	W^{0,0}_{D,D} &= -\,\Gamma_N\!\left[\varphi^{-}(\frac{U}{2},-1)
	+ \varphi^{+}(-\frac{U}{2},+1)\right],\\[6pt]
	W^{0,0}_{0,D} &=	W^{0,0}_{D,0} = W^{D,D}_{D,0} =W^{D,D}_{0,D} = +\, i\,\frac{\Gamma_S}{2},\\
	W^{0,D}_{0,0} &= W^{D,0}_{0,0} = W^{D,0}_{D,D} = W^{0,D}_{D,D} = -\, i\,\frac{\Gamma_S}{2}.
\end{align}
\label{rates_QD}
  \end{subequations}
Here $W_{a,b}\equiv W^{a,b}_{a,b}$ are the tunneling rates between diagonal elements of the reduced density matrix, and  
$f^+(\omega)=\left[1+e^{\frac{\omega-\mu_N}{T}}\right]^{-1}$,  $f^-(\omega)=1-f^{+}(\omega)$ and 
  \begin{align}
  	\varphi^{\pm}_\eta(x,\xi)= \left[f^{\pm}_\eta(x)+\xi \frac{i}{\pi}\text{Re}\left\{\psi\left(\frac{1}{2}+i\frac{x-\mu_N}{2\pi T_\eta}\right)\right\}\right]
  	\label{eq_varphi},
  \end{align}
  with $\xi\in \left\{-1,+1\right\}$, $\psi(z)$ the digamma function with complex argument $z$, and for the case of the QD, the index $\eta$ refers to the only normal lead and is omitted. From  Eq.~(\ref{rates_QD}) it is evident that the only rates that contribute explicitly to the current and the noise are the diagonal ones, since they carry the counting field dependence, although the non-diagonal terms contribute implicitly, as they are necessary to determine the reduced density matrix. 

\begin{figure}
	\centering
	\includegraphics[width=1\linewidth]{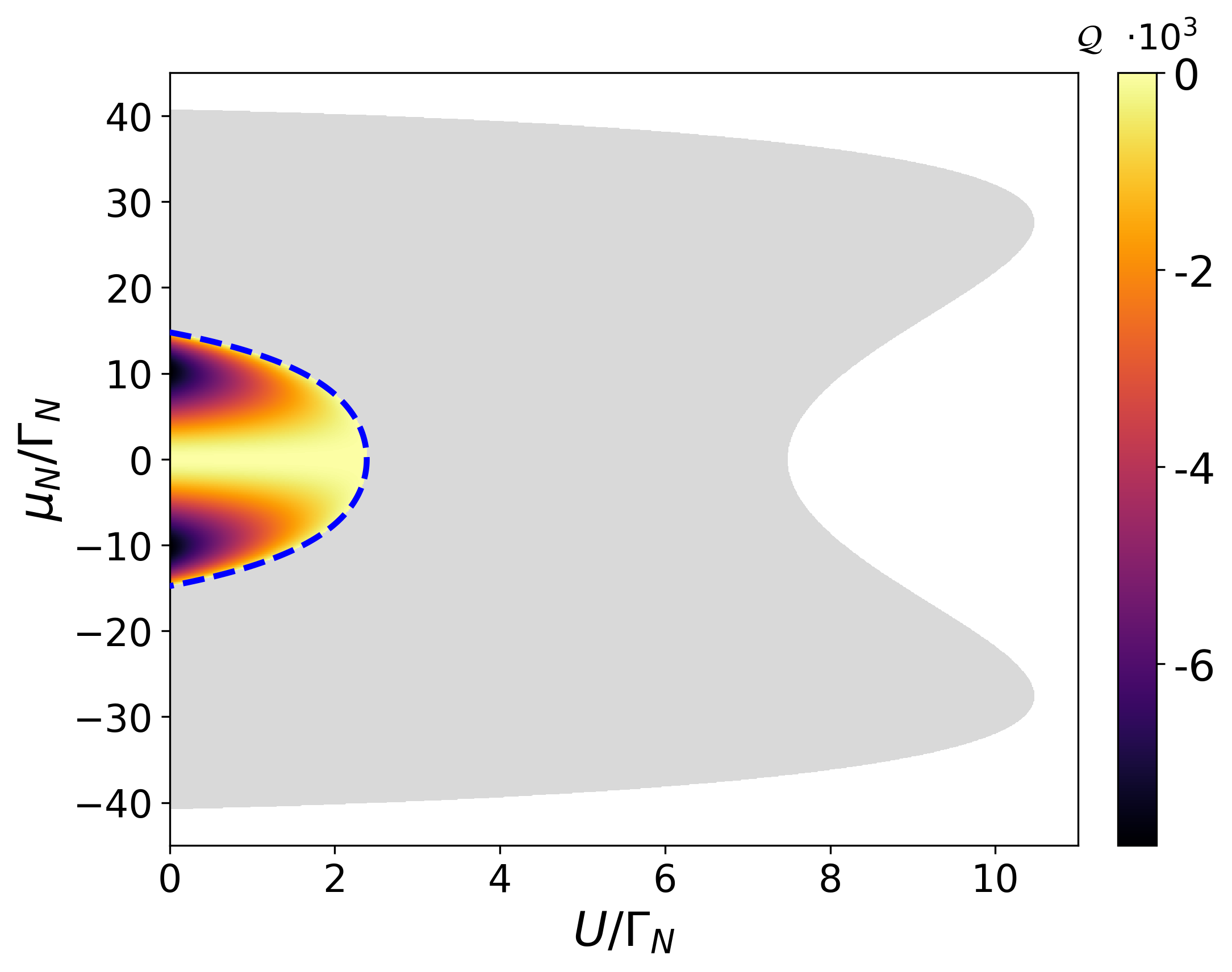}
	\caption{Violation of the quantum TUR in the single dot as a function of the Coulomb  interaction and the chemical potential of the normal lead. The dashed blue line corresponds to the saturation of the quantum bound $\mathcal{Q}=0$, and the gray area represents the region where the classical TUR is violated. The hybrid quantum bound $\mathcal{Q}_H$ is never violated. Same parameters as in Fig.~\ref{figure2}.}
	\label{figure3}
\end{figure}

In Fig.~\ref{figure2} we show the superconducting current [panel (a)], the noise [panel (b)], and the rate of entropy production  [panel (c)] as a function of the Coulomb  interaction and the chemical potential of the normal lead for $\Gamma_S=\sqrt{5/3}\Gamma_N$, $k_BT=10\Gamma_N$, at the particle-hole symmetric point  (PHSP) $\varepsilon=-U/2$.  At the PHSP, the Local Andreev Reflection (LAR) current through the dot exhibits a resonance, since $E_{{0}}=E_{{D}}$, and the current reaches its maximum value for bias voltages above the Coulomb-blockade threshold, i.e., $|\mu_N|\ge U/2$ (indicated with a dashed black lines in  Fig.~\ref{figure2}). The small finite current in the Coulomb-blockade region, that is between the dashed lines, is due to thermal smearing and vanishes as the temperature is decreased. On the other hand, the noise [panel (b)] is found to be maximum in the non-interacting equilibrium case ($U=0$ and $\mu_{N}=0$), and determined by its thermal contribution. As the interaction increases, the noise is maximized in the vicinity of the Coulomb-blockade threshold. Finally, the rate of entropy production [$\sigma=-\mu_N I/(eT)$ in our case, plotted in panel (c)] captures the combined effect of the current and the chemical potential, showing a region of small values above the Coulomb-blockade threshold for small interaction strength.


In Fig.~\ref{figure3} the corresponding quantum TUR of Eq.~(\ref{TURs}b) is shown in the region where  a violation is found. The maximum departure from the bound is obtained in the non-interacting limit  around $|\mu_N/\Gamma_N|\sim10$ and it evolves smoothly as a function of $U$ until the inequality is restored at $U\sim 2\Gamma_N$.   The blue dashed line corresponds to the saturation of the quantum TUR.  The gray area in Fig.~\ref{figure3} represents the region in which a violation of the classical TUR is observed. No violation is observed for the hybrid quantum TUR [Eq.~(\ref{TURs}c)]  and the hybrid quantum bound is saturated in the $\mu_N\to 0$ limit.  When the ratio between the coupling strengths $\Gamma_S/\Gamma_N$ is increased, the TUR violation is rapidly suppressed (not shown).

\begin{figure}
	\centering
	\includegraphics[width=1\linewidth]{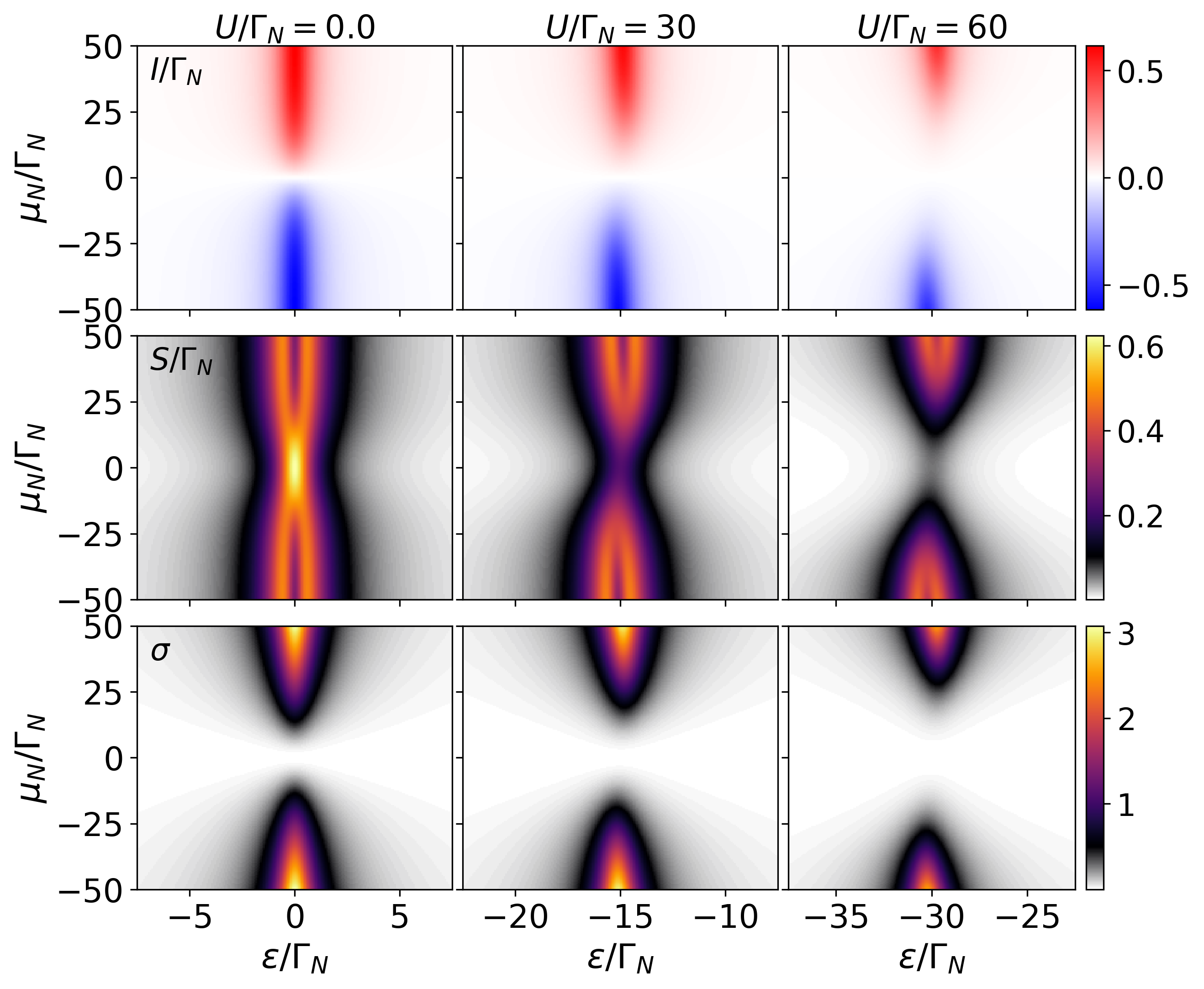}
	\caption{ (Up) Superconducting current, (Center) Noise, and (Down) Rate of entropy production in the single quantum dot as a function of the gate level $\varepsilon$ and the chemical potential of the normal lead. Each column corresponds to a different Coulomb interaction, $U/\Gamma_N=0,30,60$ from left to right. The rest of parameters are $\Gamma_S=\sqrt{5/3}\Gamma_N$, and $k_BT=10\Gamma_N$.
	}
	\label{figure4}
\end{figure}

\begin{figure}
	\centering
	\includegraphics[width=1\linewidth]{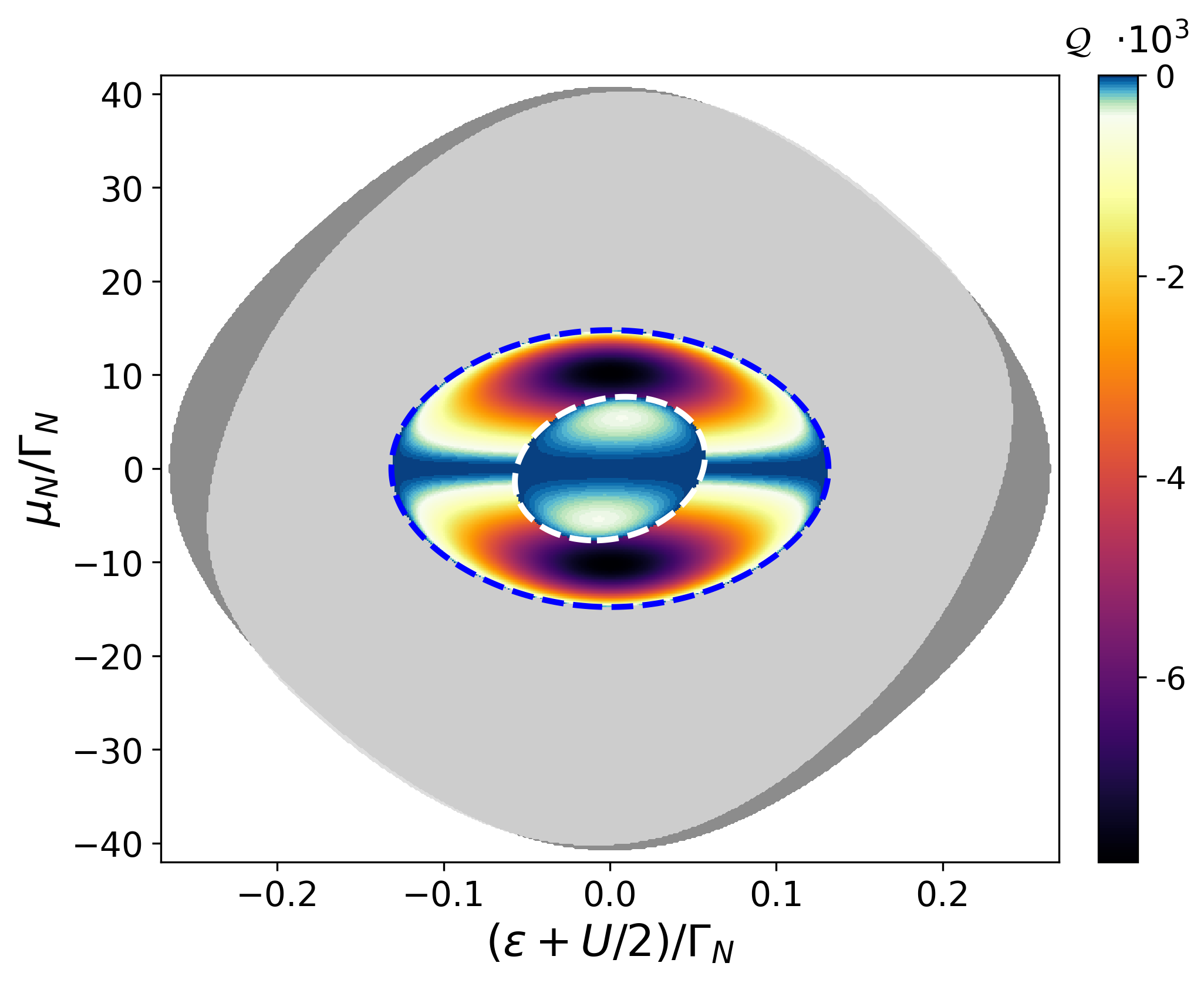}
	\caption{Violation of the quantum TUR in the single dot as a function of the gate level $\varepsilon$, centered at the LAR resonance condition $\varepsilon+U/2=0$, and the chemical potential of the normal lead. The outer colored region corresponds to $U=0$. The inner colored region corresponds to $U/\Gamma_N=2$. The dashed blue and white lines correspond to the saturation of the quantum bound $\mathcal{Q}=0$, for  $U/\Gamma_N=0,2$, respectively. The dark and light gray areas represents the regions where the classical TUR is violated for  $U/\Gamma_N=0,2$, respectively. Same parameters as in Fig.~\ref{figure4}.}
	\label{figure5}
\end{figure}
Fig.~\ref{figure4} shows the superconducting current, noise, and rate of entropy production 
as a function of the gate level $\varepsilon$ and the chemical potential $\mu_N$ of the normal lead, for different values of the interaction strength $U/\Gamma_N$. In the noninteracting case ($U=0$), the current exhibits a narrow vertical stripe centered at the  energy level value corresponding to the LAR resonance. As the interaction is increased, this structure evolves opening a gap centered around $\mu_N \simeq 0$, reflecting the onset of Coulomb blockade smeared out by the thermal broadening. In the low temperature limit, the current develops asymmetric tails around $|\mu_N| \sim U/2$, leading to a weak curvature of the resonant feature in the $(\varepsilon,\mu_N)$ plane, as observed in Ref.~\cite{pala2007nonequilibrium}. This “nose”-like feature originates from interaction-induced level-renormalization effects associated with coherent processes. Although these effects are suppressed at the relatively high temperature considered here ($k_B T = 10\Gamma_N$), a small but systematic shift of the resonance remains visible: the current maximum is displaced toward smaller (larger) energy level values for $\mu_N < 0$ ($\mu_N > 0$).
 The noise shows a qualitatively similar evolution. With increasing interaction strength, a gap proportional to $U$ opens around $\mu_N \simeq 0$, 
 and the resonance is progressively shifted by the level-renormalization effects.
 The noise reaches its largest values in the noninteracting case near $\mu_N \sim 0$ and is gradually suppressed as $U$ increases. In contrast to the current, however, the noise exhibits a pronounced narrow dip along the resonant condition, where its value is significantly reduced. The rate of entropy production follows the same overall trend. Increasing the interaction strength opens a Coulomb-induced gap and leads to a reduction of the entropy production rate at resonance.

For a small but finite Coulomb interaction $U/\Gamma_N = 2$,  the current, noise, and rate of entropy production look the same as the non-interacting results shown in the left column of Fig.~\ref{figure4}. 
Due to the relatively high temperature considered here, 
the Coulomb-blockade features that are prominent at low temperatures are significantly smeared out, making differences between the two cases difficult to resolve at the level of these observables. By contrast, the quantum TUR proves to be a much more sensitive quantity and it captures these differences. This is illustrated in Fig.~\ref{figure5}, where the violation of the quantum TUR is quantified for $U/\Gamma_N = 0$ and $U/\Gamma_N = 2$ as a function of the gate level $\varepsilon$ and the chemical potential. The outer colored region corresponds to the non-interacting case, while the inner region represents the interacting case with $U/\Gamma_N = 2$. The dashed blue and white lines indicate saturation of the quantum bound, $\mathcal{Q}=0$, for $U/\Gamma_N = 0$ and $U/\Gamma_N = 2$, respectively. The dark- and light-gray areas denote the regions where the classical TUR is violated for $U/\Gamma_N = 0$ and $U/\Gamma_N = 2$. It is observed that the largest quantum-TUR violations occur in the noninteracting case, for gate levels close to resonance and for chemical potentials $|\mu_N/\Gamma_N |\sim 10$. In the interacting case, $U/\Gamma_N = 2$, the violation remains finite but is restricted to a smaller region of parameter space and is substantially reduced in magnitude. Also in this case the hybrid quantum TUR [Eq.~(\ref{TURs}c)] is never violated.

\section{Cooper Pair Splitter}
\label{Sectionfour}
To extend our analysis beyond local Andreev transport, we consider a Cooper-pair splitter (CPS), which provides a minimal platform to study the interplay between Coulomb interactions, nonlocal superconducting coherence, and current precision. The central region consists of two interacting quantum dots described by the Hamiltonian
\begin{align*}
	H_{DQD}=\sum_{\alpha=L,R}\left(\varepsilon_\alpha \sum_{\sigma}n_{\alpha\sigma}+U_\alpha n_{\alpha \uparrow},n_{\alpha\downarrow}\right)+U_{LR}n_1n_2.
\end{align*}
The central region is coupled to two normal leads $\alpha=L,R$ at equal temperatures $T$ and chemical potentials $\mu_N$, and as in the previous model, and to one superconducting lead grounded with $\Delta\to \infty$, as schematically illustrated in Fig.~\ref{figure1} b). The Hamiltonian of the full CPS system is 
 \begin{align}
 	H&= H_{DQD} - \frac{\Gamma_{\rm S}}{2}\sum_{\eta}\,\left(d^\dagger_{\eta,\uparrow} d^\dagger_{\eta,\downarrow}+ {\rm H.c.} \right)\nonumber\\&- \frac{\Gamma_{\rm C}}{2}\,\left(d^\dagger_{R,\uparrow} d^\dagger_{L,\downarrow} -d^\dagger_{R,\downarrow} d^\dagger_{L,\uparrow} + {\rm H.c.} \right)
 	+H_{\rm leads}+H_{{\rm tunn}}\,.
 		\label{H_CPS}
 \end{align}
 For simplicity, in the following we assume $U_L=U_R\equiv U$.
The second term in Eq.~(\ref{H_CPS}) corresponds to the LAR contribution, while the third term describes Cross (nonlocal) Andreev Reflection (CAR).

Following a procedure similar to the single-dot case, we now compute the rates $W^{\xi_1\xi_1'}_{\xi_2\xi_2'}$ to first order not only in $\Gamma_N$, but also in $\Gamma_S$, and $\Gamma_C$. This approximation allows us to accurately characterize the reduced density matrix for any gate level, but for values of the chemical potential large enough such that higher-order processes not accounted here do not play a major role. We therefore restrict the couplings such that $k_BT\gg \Gamma_N, \Gamma_S,\Gamma_C$.

Since the basis of the Fock space of the double dot consists of 16 states, the number of coherences that contribute to the rate matrix is $240$.  We have derived an algorithm to evaluate numerically the elements of the rate matrix based on the general structure of the different diagrams that contribute to them. The details on the derivation can be found in Appendix~\ref{App1}. 
It is worth noting that for $\Gamma_S=\Gamma_C$ the number of finite coherences contributing to the master equation is 50  and reduces to only 20 when either $\Gamma_S$ or $\Gamma_C$ is zero.
	\begin{figure}
	\centering
	\includegraphics[width=1\linewidth]{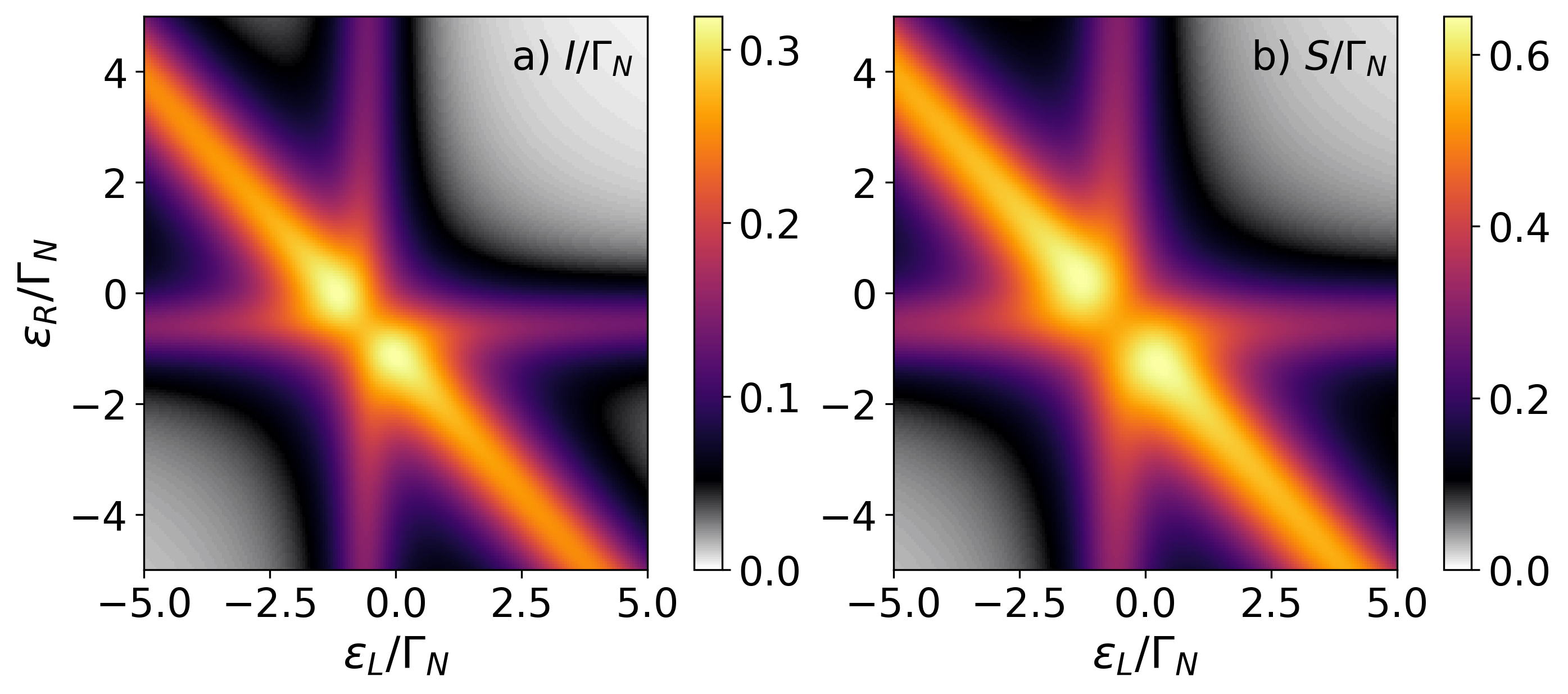}
	\caption{(a) Superconducting current, and (b) noise as a function of the gate levels $\varepsilon_{L}$, and $\varepsilon_{R}$ in the CPS. The parameters are  $U=U_{LR}=0.3\Gamma_N$,  $\Gamma_S=\Gamma_C=\sqrt{5/12}\Gamma_N$, and $\mu_N=k_B T=10\Gamma_N$.}
	\label{figure6}
\end{figure}

In Fig.~\ref{figure6} we present the superconducting current and zero-frequency noise of the CPS system as functions of the gate levels $\varepsilon_L$ and $\varepsilon_R$, for $U = U_{LR} = 0.3\,\Gamma_N$, $\Gamma_S = \Gamma_C = \sqrt{5/12}\,\Gamma_N$, and $\mu_N = k_B T = 10\,\Gamma_N$. Both the current and the noise exhibit three distinct resonances, which originate from LAR processes in the left and right quantum dots, as well as from CAR. The LAR resonance associated with the left dot occurs when the many-body states 
$\ket{0,\chi_R}$ and $\ket{\uparrow\downarrow,\chi_R}$, where $\chi_R$ denotes a state of the right dot, become degenerate. 
This condition is achieved when the left-dot level satisfies $\varepsilon_L = -U/2 - U_{LR} n_R$, where $n_R$ is the occupation of the right dot in the state $\chi_R$. Analogously, the LAR resonance of the right dot corresponds to the degeneracy between the states 
$\ket{\chi_L,0}$ and $\ket{\chi_L,\uparrow\downarrow}$, which is obtained for $\varepsilon_R = -U/2 - U_{LR} n_L$. 
The CAR resonance occurs when the energies of the states $\ket{\chi_L,\chi_R}$ and $\ket{\chi_L \sigma,\chi_R \bar{\sigma}}$ coincide, where $\chi_L\in\{0,\bar\sigma\}$ $\chi_R\in\{0,\sigma\}$. 
This condition can be written as $\varepsilon_L + \varepsilon_R + U (n_L + n_R) + U_{LR} (n_L + n_R + 1) = 0$, where $n_L$ and $n_R$ are the occupations corresponding to the states $\chi_L$ and $\chi_R$, respectively.  As a result, both the current and the noise display an enhanced signal along the diagonal line in the $(\varepsilon_L,\varepsilon_R)$ plane associated with the CAR process, with particularly pronounced features close to the PHSP. Compared to the current, the noise exhibits broader resonant structures, and the corresponding Fano factor in the vicinity of the maxima is approximately equal to two.
\begin{figure}
	\centering
	\includegraphics[width=1\linewidth]{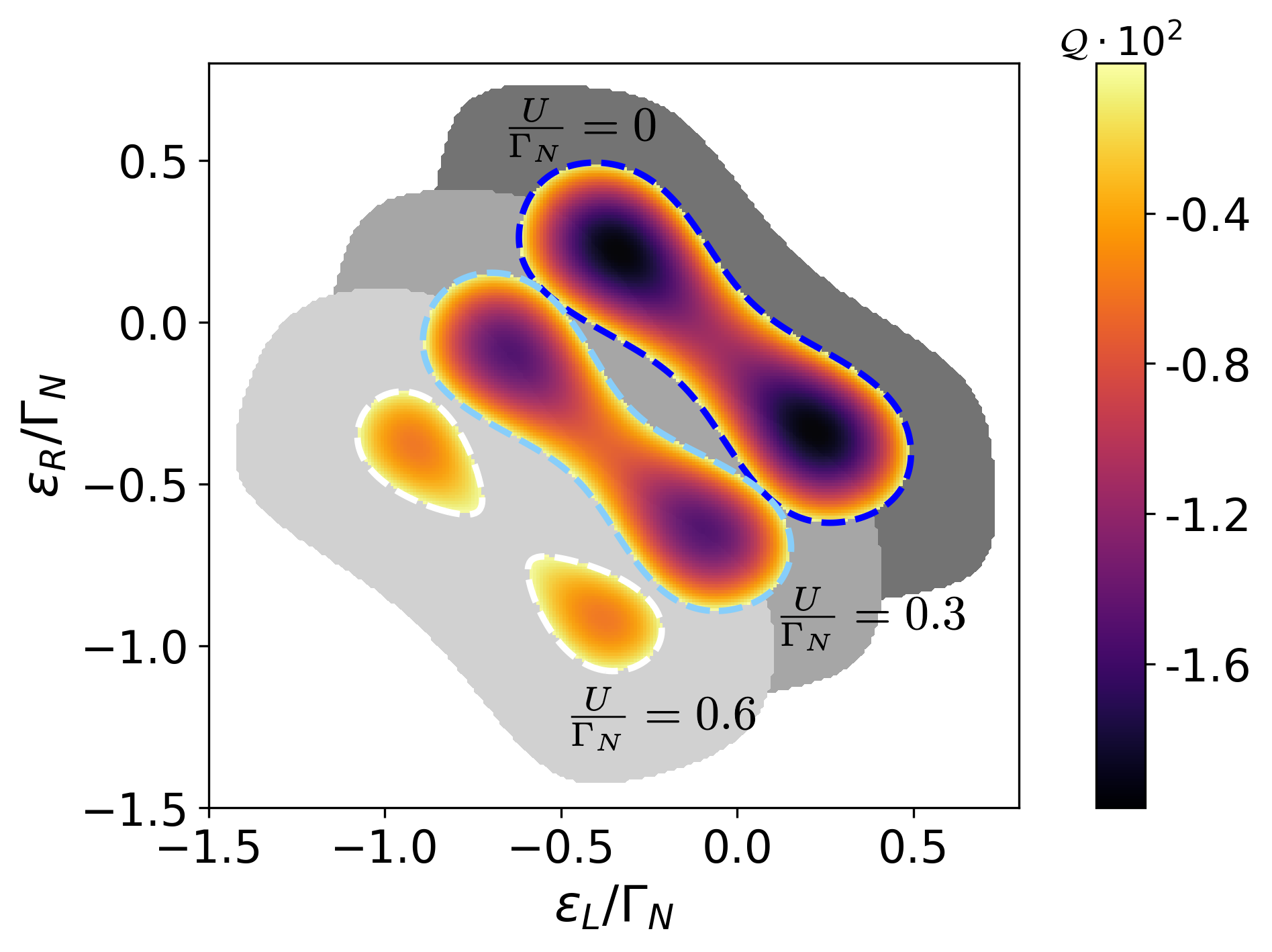}
	\caption{Violation of the quantum TUR in the CPS as a function of the gate levels $\varepsilon_{L}$, and $\varepsilon_{R}$, for three values of the interaction $U/\Gamma_N=U_{LR}/\Gamma_N=0,0.3,0.6$. As the interaction increases, the area of the region where the quantum TUR is violated is decreased and evolves to more lower values of the gates. The dashed dark blue, light blue and white lines correspond to the saturation of the quantum bound $\mathcal{Q}=0$, for  $U/\Gamma_N=0,0.3,0.6$, respectively. The gray areas represents the regions where the classical TUR is violated for  $U/\Gamma_N=0,0.3,0.6$, from dark to light. Same parameters as in Fig.~\ref{figure6}.}
	\label{figure7}
\end{figure}
In Fig.~\ref{figure7} we show the violation of the quantum TUR in the CPS as a function of the gate levels $\varepsilon_L$ and $\varepsilon_R$, for $U/\Gamma_N = 0,\,0.3,0.6$, and for the same parameters as in Fig.~\ref{figure6}. The strongest departure from the quantum bound is observed in the noninteracting case, where the violation is maximal along the CAR resonance, in agreement with previous findings~\cite{mayo2025thermodynamic}.  As the interaction strength is increased, the region in parameter space where the quantum TUR is violated progressively shrinks and splits into two disconnected lobes. These regions are shifted toward lower gate values and closely follow the interaction-induced evolution of the CAR contribution observed in the current and noise. At the same time, the magnitude of the violation is substantially reduced, indicating a progressive suppression of macroscopic quantum coherence by Coulomb interactions. 
The dashed dark-blue, light-blue, and white contours correspond to the saturation of the quantum bound, $\mathcal{Q}=0$, for $U/\Gamma_N = 0,\,0.3,$ and $0.6$, respectively. The gray shaded areas indicate the regions where the classical TUR is violated, with darker to lighter shading corresponding to increasing interaction strength.

\begin{figure}
	\centering
	\includegraphics[width=1\linewidth]{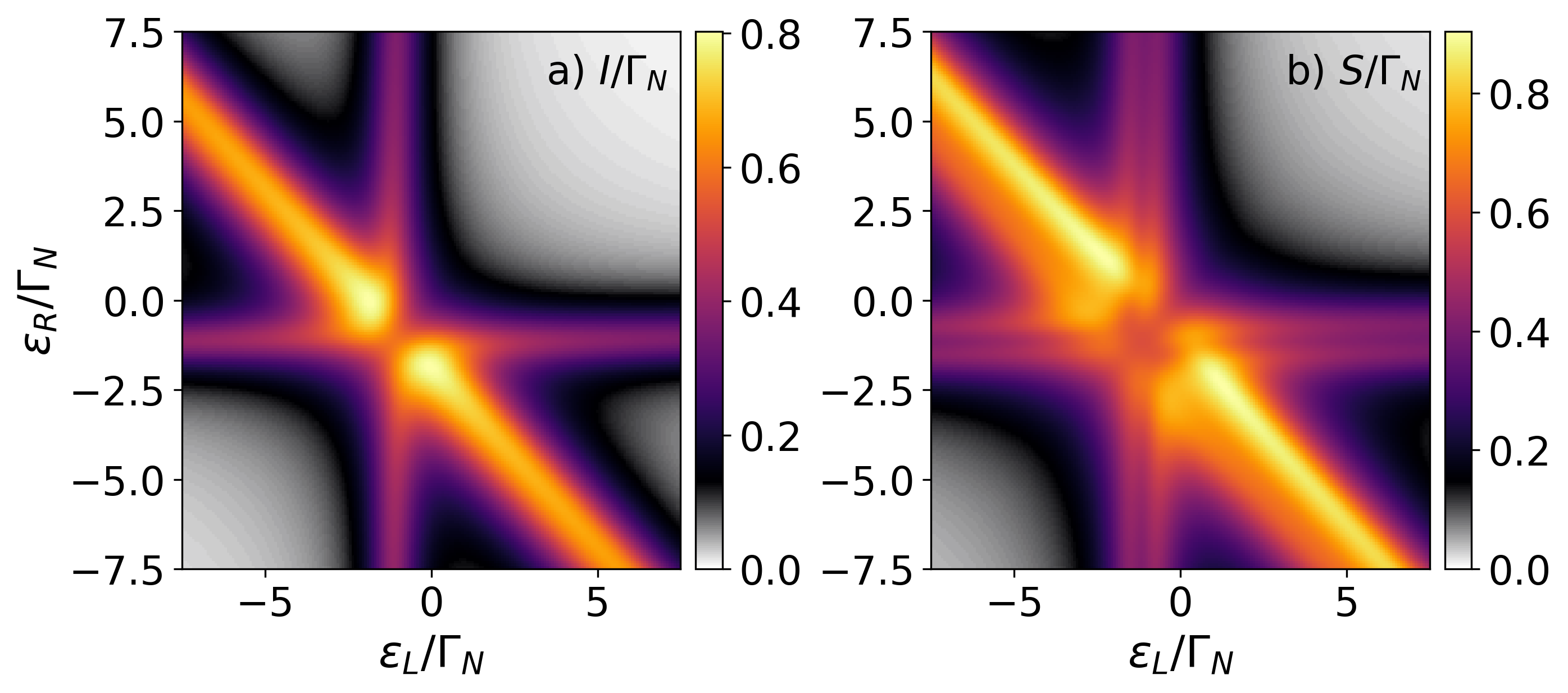}
	\caption{(a) Superconducting current, and (b) noise as a function of the gate levels $\varepsilon_{L}$, and $\varepsilon_{R}$ in a CPS system. The parameters are  $U=U_{LR}=0.5\Gamma_N$ and $\Gamma_S=\Gamma_C=\Gamma_N$, and $\mu_N= 2 k_B T=20\Gamma_N$.}
	\label{figure8}
\end{figure}

\begin{figure}
	\centering
	\includegraphics[width=\linewidth]{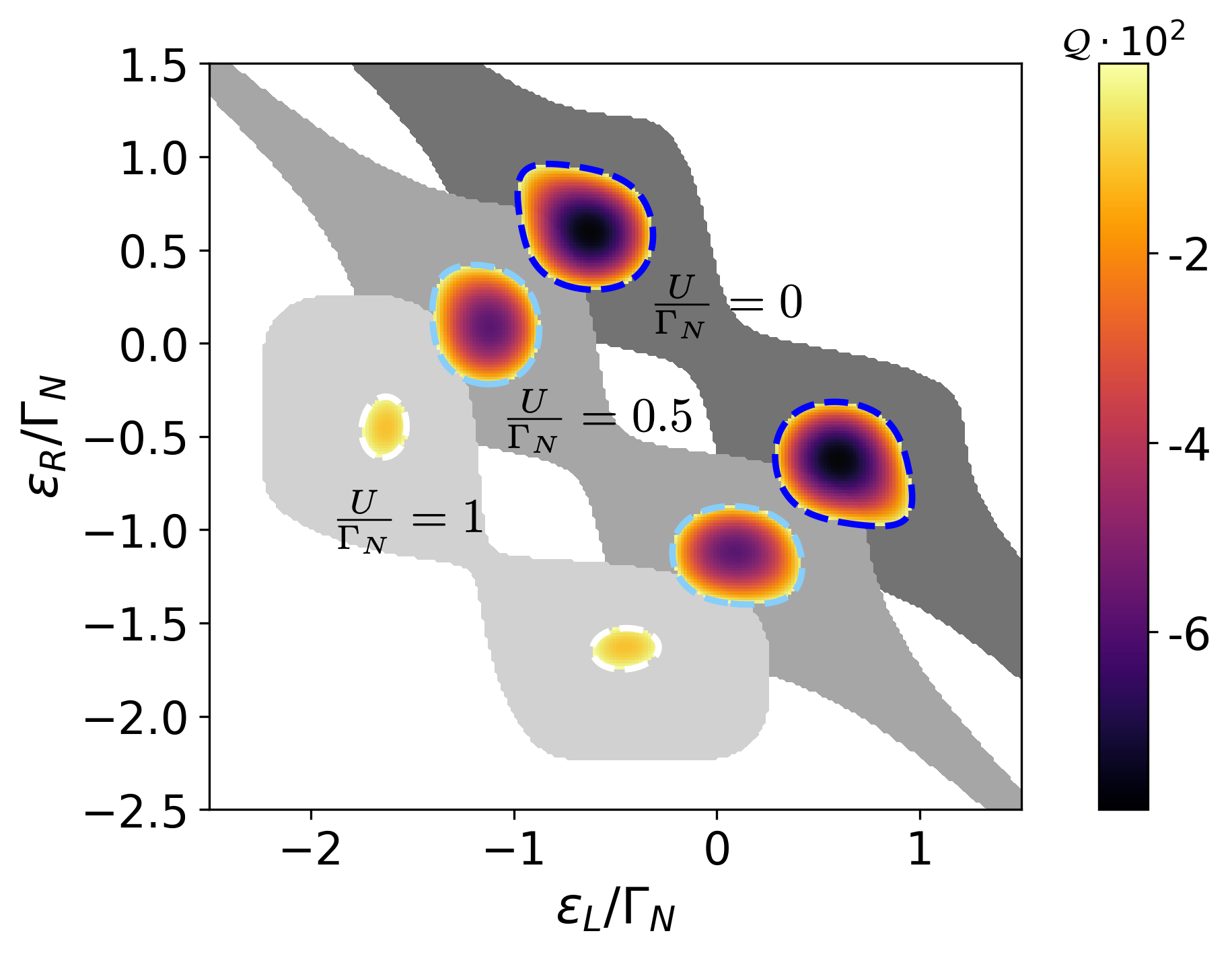}
	\caption{Violation of the quantum TUR in the CPS as a function of the gate levels $\varepsilon_{L}$, and $\varepsilon_{R}$, for three values of the interaction $U/\Gamma_N=U_{LR}/\Gamma_N=0,0.5,1$. As the interaction increases, the area of the region where the quantum TUR is violated is decreased and evolves to more lower values of the gates. The dashed dark blue, light blue and white lines correspond to the saturation of the quantum bound $\mathcal{Q}=0$, for  $U/\Gamma_N=0,0.5,1$, respectively. The gray areas represents the regions where the classical TUR is violated for  $U/\Gamma_N=0,0.5,1$, from dark to light. Same parameters as in Fig.~\ref{figure8}.}
	\label{figure9}
\end{figure}

In Fig.~\ref{figure8} we present the superconducting current and zero-frequency noise of the CPS as functions of the gate levels $\varepsilon_L$ and $\varepsilon_R$, now considering equal coupling strengths $\Gamma_N=\Gamma_S=\Gamma_C$, a larger interaction $U=U_{LR}=0.5\,\Gamma_N$, and a higher bias $\mu_N = 2 k_B T = 20\,\Gamma_N$. The overall structure of both observables closely resembles that found in Fig.~\ref{figure6}. However, due to the modified local occupations induced by the stronger interaction and symmetric couplings, the resonant Andreev-reflection processes occur at shifted gate values. As a consequence, the resonant features become broader, and a clear splitting of the maxima is visible in the noise profile, while the current retains a smoother structure. The corresponding violations of the quantum TUR are shown in Fig.~\ref{figure9} for  $U=U_{LR}=0,\,0.5,1\,\Gamma_N$. In the non-interacting case, the maximum violation is enhanced by nearly a factor of five compared to the parameter regime of Fig.~\ref{figure7}, and it is localized in two distinct regions around the CAR resonance. As the interaction strength increases, the magnitude of the violation is progressively reduced but remains finite up to $U=U_{LR}\sim\Gamma_N$, indicating a persistence of quantum-coherent transport effects. The dashed dark-blue, light-blue, and white contours denote saturation of the quantum bound, $\mathcal{Q}=0$, for $U/\Gamma_N = 0,\,0.5,$ and $1$, respectively. The gray shaded regions indicate violations of the classical TUR, with darker to lighter shading corresponding to increasing interaction strength. The hybrid quantum TUR is never violated in this situation.

\subsection*{Limit $U_i \to \infty$}

We now turn to the limit $U_i \to \infty$, which realizes the regime where double occupancy of each quantum dot is completely forbidden,  leading to a substantial reduction in the number of many-body states and coherent contributions entering the rate matrix. This limit completely suppresses local Andreev processes and constrains superconducting transport to be mediated exclusively by nonlocal correlations. An analogous suppression of local pairing coherence can also be achieved by applying a sufficiently strong Zeeman field, which energetically separates spin-resolved levels and inhibits the formation of local singlet pairs. The $U_i \to \infty$ limit therefore provides a stringent benchmark to assess TURs in regimes where local superconducting coherence is absent. 

In this limit, the CPS admits nine accessible basis states, $
\left\{ \ket{0}, \ket{\sigma,0}=c_{L\sigma}^\dagger\ket{0},\ket{0,\sigma}=c_{R\sigma}^\dagger\ket{0}, \ket{\sigma,\sigma'}=c_{L\sigma}^\dagger c_{R\sigma'}^\dagger\ket{0}\right\},
$
where $\sigma,\sigma' \in \{\uparrow,\downarrow\}$. The two-electron states are degenerate and share the energy $E = \varepsilon_L + \varepsilon_R + U_{LR}$.

As in the finite-$U_i$ case, we compute the transition rates $W^{\xi_1\xi_1'}_{\xi_2\xi_2'}$ to first order in both $\Gamma_N$ and $\Gamma_C$, and at  $\delta = 0$, corresponding to the energy difference between the empty and $\ket{\sigma\bar\sigma}$ states. Owing to the reduced size of the Fock space, the dimension of the rate matrix is now considerably smaller, namely $15 \times 15$. Only coherence terms connecting the empty state $\ket{0}$ with the two-electron states $\ket{\uparrow\downarrow}$ and $\ket{\downarrow\uparrow}$, as well as the coherences between $\ket{\uparrow\downarrow}$ and $\ket{\downarrow\uparrow}$ themselves, need to be retained.

A detailed derivation of all generalized rates in this limit is provided in Appendix~\ref{App1}. 
\begin{figure*}
	\centering
	\includegraphics[width=0.8\linewidth]{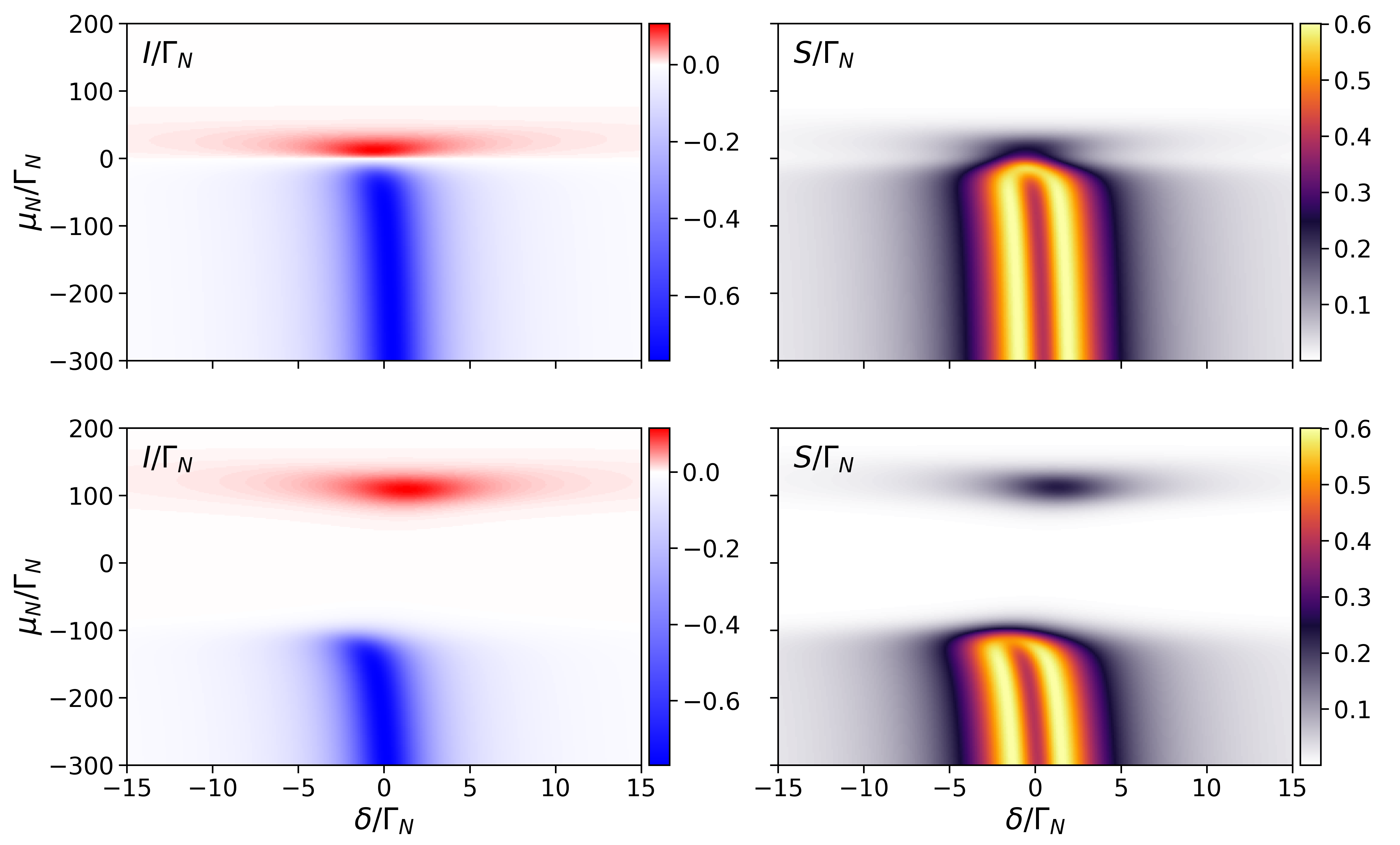}
	\caption{Superconducting current $I/\Gamma_N$, and noise $S/\Gamma_N$ of the CPS as a function of the detuning $\delta=\varepsilon_L+\varepsilon_R+U_{LR}$ and the chemical potential $\mu_N$. The upper and lower panels correspond to $U_{LR}/\Gamma_N=0,200$, respectively. The rest of the parameters are $k_BT=10\Gamma_N$ and $\Gamma_C=\sqrt{5/3}\Gamma_N$.}
	\label{figure10}
\end{figure*}

In Fig.~\ref{figure10} we present the current and the zero-frequency noise associated with the superconducting lead of the CPS as functions of the chemical potential $\mu_N$ and the detuning $\delta$, for $k_B T = 10\,\Gamma_N$ and $\Gamma_C = \sqrt{5/3}\,\Gamma_N$. The upper panels correspond to the $U_{LR}=0$ case, while the lower panels show the regime of strong interdot repulsion, $U_{LR}=200\Gamma_N$.

Despite the different interaction strengths, the qualitative transport behavior is similar in both cases. For $\mu_N < -U_{LR}/2$, Cooper pairs originating from the superconducting lead are split between the two dots and subsequently emitted into the normal leads. This process is enabled by the near degeneracy of the states $\ket{\eta\sigma}$, $\ket{\uparrow\downarrow}$, and $\ket{\downarrow\uparrow}$ around $\delta = 0$, which gives rise to the pronounced current and noise features observed in this region.  

In the intermediate regime $-U_{LR}/2 < \mu_N < U_{LR}/2$, the double dot is predominantly occupied by a single electron. As a consequence, transport through the superconducting lead is suppressed, since additional electrons cannot enter or leave the system via Andreev processes. 
For $\mu_N \gg U_{LR}/2$, the  system is stuck in a triplet configuration and decouples from the superconductor. This is the so called triplet blockade \cite{pala2007nonequilibrium}. However, there is a region of width $k_B T$ around $\mu_N \gtrsim U_{LR}/2$, where the current is non vanishing due to the temperature broadening of the Fermi functions.

We have analyzed both the classical and quantum TURs in this parameter regime and find no violation of either bound. Only a saturation of the bounds is observed in the limit $\mu_N \to 0$, while small values of the TURs parameters in Eq.~(\ref{TURs}) are also reached within the transport region around $\delta \sim 0$ for $\mu_N < -U_{LR}/2$.
This behavior is consistent with the results of the previous section (see Fig.~\ref{figure8}), which establish a reduction of the violation of both TURs with increasing values of $U$.
It might be tempting to try to understand the behavior of current and noise in the upper panels of Fig.~\ref{figure10} (relative to $U_{LR}=0$) using a mapping with the case of a non-interacting single quantum dot of Sec.~\ref{Sectionthree} (see the first column of Fig.~\ref{figure4}).
Apparently, the only difference between the two cases is that the two electrons provided by the superconductor are transferred locally (for the single quantum dot) or strictly non-locally (for the CPS).
Such mapping, however, does not work since, in the CPS case, interaction indeed plays a role by inducing correlations that increase the number of possible states involved in the transport.
As a result, current and, especially, noise are different (but not too much) in the two cases, leading to the fact that neither the classical nor the quantum TUR is violated for the CPS.

\section{Conclusions}
\label{Sectionfive}
In this work we have studied the nonequilibrium Andreev-mediated transport and current precision in interacting hybrid normal--superconducting systems. Using the generalised master equation combined with real-time diagrammatics and full counting statistics, we computed the steady-state current, zero-frequency noise, and rate of entropy production in quantum-dot--based systems in the large superconducting-gap limit. Treating Coulomb interactions exactly within a reduced density-matrix formalism, we systematically assess how interactions modify transport resonances,
and ultimately constrain current precision.

Our results show that Coulomb interactions have a pronounced impact on superconducting transport well beyond their effect on average currents. While Andreev processes continue to dominate subgap transport, interactions renormalize the relevant resonant conditions, suppress coherent contributions, and progressively reduce current precision. These effects are particularly visible in regimes of strong thermal broadening, where conventional Coulomb-blockade signatures are largely washed out but  TURs remain highly sensitive.

By analyzing the TURs, we showed that precision bounds provide a quantitative way to characterize how interactions constrain Andreev-mediated transport beyond what is visible at the level of average currents. In the noninteracting regime, violations of the quantum TUR arise in parameter regions associated with resonant Andreev transport. In the Cooper-pair splitter, the strongest departures from the quantum bound are found along the cross Andreev reflection resonance, while the overall structure of the violation regions reflects the combined influence of local and nonlocal Andreev processes. As Coulomb interactions increase, these violations are progressively reduced in both magnitude and extent, and are eventually suppressed once interaction strengths become comparable to the tunnel couplings. In contrast, the hybrid quantum bound remains satisfied throughout and is only approached in the vanishing-current limit, confirming its robustness for Andreev-dominated transport.

Overall, our findings establish current precision as a powerful probe of interaction-induced renormalization and decoherence in hybrid superconducting devices. They clarify how superconducting coherence, electron--electron interactions, and nonequilibrium fluctuations jointly constrain transport far from equilibrium, and provide a systematic framework to assess the performance of interacting superconducting systems.

\section*{Acknowledgments}
We acknowledge support from ANPCyT (PICT-2019-04349 and PICT-2021-01288) (F.M.), the European Union under the Horizon Europe research and innovation programme (Marie Skłodowska-Curie grant agreement no. 101148213, EATTS) (N.S.), MUR-PRIN 2022 - Grant No. 2022B9P8LN - (PE3)-Project NEThEQS “Non-equilibrium coherent
thermal effects in quantum systems” in PNRR Mission 4 - Component 2 - Investment 1.1 “Fondo per il Programma Nazionale di Ricerca e Progetti di Rilevante Interesse
Nazionale (PRIN)” (F.T.), PNRR MUR project
PE0000023-NQSTI (R.F), the Royal Society through the International Exchanges between the UK and Italy (Grants No.
IEC R2 192166) (F.T), and the European Union (ERC, RAVE, 101053159) (R.F). 

\appendix
\section{Calculation of the  rates from real time diagrammatics}\label{App1}
\label{Appendixone}

\subsection{Cooper pair splitter: Finite $U_\alpha$ and $U_{LR}$}

In this section we derive a general and computationally efficient representation of the rate-matrix elements $W$ within the real-time diagrammatic approach. The rate matrix can be systematically decomposed into five distinct classes of contributions. These include: (i) diagonal population rates, $W_{A} \equiv W^{A,A}_{A,A}$; (ii) off-diagonal population rates, $W_{A,B} \equiv W^{A,B}_{A,B}$; (iii) diagonal coherent rates, $W^{A}_{B} \equiv W^{A,A}_{B,B}$; (iv) off-diagonal coherent rates of the form $W^{A,C}_{B,D}$ that are finite only when either $A=C$ or $B=D$; and (v) fully off-diagonal coherent rates of the form $W^{A,C}_{B,D}$, with $A \neq B \neq C \neq D$. Each class of rates admits a compact expression that can be evaluated efficiently for numerical calculations.
 
 For simplicity in the following we use $\int\equiv\frac{1}{2\pi}\int d\omega$ and $\omega^+\equiv \omega+i0^+$
\subsubsection*{1 -- Diagonal population rates $W_{A} \equiv W^{A,A}_{A,A}$}
The diagonal population rates describe the total probability loss from a many-body state $\ket{A}$ due to tunneling events involving the normal leads. They are given by
\small
\begin{align*}
	W_{A} = -\Gamma_N \left( 
	\sum_{(\eta\sigma)_{\mathrm{empty}}} f_\eta^{+}(E_{+}-E_A)
	+ \sum_{(\eta\sigma)_{\mathrm{occ}}} f_\eta^{-}(E_A-E_{-})
	\right),
\end{align*}
\normalsize
where the first sum runs over all spin--orbital states $(\eta\sigma)$ that are unoccupied in $\ket{A}$ and for which electron addition is allowed, i.e., $c^\dagger_{\eta\sigma}\ket{A} \neq 0$. The second sum runs over all $(\eta\sigma)$ states that are occupied in $\ket{A}$ and for which electron removal is allowed, i.e., $c_{\eta\sigma}\ket{A} \neq 0$. Here, $E_{+}$ ($E_{-}$) denotes the energy of the many-body state obtained after adding (removing) one electron with spin $\sigma$ in dot $\eta$ to (from) the state $\ket{A}$ with energy $E_A$.

As an example we explicitly compute the rate 
\small
\begin{align*}
	&W_{(\uparrow,\downarrow)}=
	\RTDthree{}{(\uparrow,\ud)}{}{R\uparrow}{}{}{<-}+
	\RTDthree{}{(\ud,\downarrow)}{}{L\downarrow}{}{}{<-}+\RTDfour{}{}{R\uparrow}{}{(\uparrow,\ud)}{}{->}\\&+
	\RTDfour{}{}{L\downarrow}{}{(\ud,\downarrow)}{}{->}
    +\RTDthree{}{L\uparrow}{}{R\downarrow}{}{}{->}+
	\RTDthree{}{R\downarrow}{}{L\uparrow}{}{}{->}\\&+\RTDfour{}{}{R\downarrow}{}{L\uparrow}{\ud}{<-}+
	\RTDfour{}{}{L\uparrow}{}{R\downarrow}{}{<-}
	\nonumber\\
	&= -i\Gamma_N   \left(
	\int \frac{f^+_{R}(\omega)}{\omega^+-(E_{(\uparrow,\ud)}-E_{(\uparrow,\downarrow)})}+
	\int \frac{f^+_{L}(\omega)}{\omega^+-(E_{(\ud,\downarrow)}-E_{(\uparrow,\downarrow)})}\right.\\&\left.+	\int \frac{f^+_{R}(\omega)}{(E_{(\uparrow,\ud)}-E_{(\uparrow,\downarrow)})-\omega^+}+
	\int \frac{f^+_{L}(\omega)}{(E_{(\ud,\downarrow)}-E_{(\uparrow,\downarrow)})-\omega^+}\right)\nonumber\\
	& \quad-i\Gamma_N\left(\int \frac{f^-_{R}(\omega)}{(E_{(\uparrow,\downarrow)}-\varepsilon_L)-\omega^+}+
	\int \frac{f^-_{L}(\omega)}{(E_{(\uparrow,\downarrow)}-\varepsilon_R)-\omega^+}\right.\\&\left.+\int \frac{f^-_{R}(\omega)}{\omega^+-(E_{(\uparrow,\downarrow)}-\varepsilon_L)}+
	\int \frac{f^-_{L}(\omega)}{\omega^+-(E-\varepsilon_R)}
	\right)\nonumber\\
	&=-\Gamma_N \left( f^+_R(E_{(\uparrow,\ud)}-E_{(\uparrow,\downarrow)}) +f^+_L(E_{(\ud,\downarrow)}-E_{(\uparrow,\downarrow)})\right.\\&\left.
	+f^-_R(E_{(\uparrow,\downarrow)}-\varepsilon_L) +f^-_L(E_{(\uparrow,\downarrow)}-\varepsilon_R)
	\right)
\end{align*}
\normalsize
\subsubsection*{2 -- Off-diagonal population rates $W_{A,B} \equiv W^{A,B}_{A,B}$}

The off-diagonal population rates describe tunneling-induced transitions between distinct many-body states $\ket{A}$ and $\ket{B}$. They are given by
\begin{subequations}
	\begin{align*}
		W_{A,B}(N_A < N_B) &= \Gamma_N \, \big| \langle B | c^\dagger_{\eta\sigma} | A \rangle \big| \,
		f^-_\eta(E_B - E_A)\, e^{i\chi}, \\
		W_{A,B}(N_A > N_B) &= \Gamma_N \, \big| \langle A | c^\dagger_{\eta\sigma} | B \rangle \big| \,
		f^+_\eta(E_A - E_B)\, e^{-i\chi},
	\end{align*}
\end{subequations}
for any spin--orbital index $(\eta\sigma)$. In both cases, a finite contribution arises only when the two states differ by exactly one electron, i.e., $|N_A - N_B| = 1$, while all remaining electrons occupy identical orbitals with the same spin configuration in both states.

As an example we explicitly compute the rates
\small
\begin{align*}
	&W_{L\uparrow,(\uparrow,\downarrow)}=	\RTDtwo{(\uparrow,\downarrow)}{L\uparrow}{R\downarrow}{(\uparrow,\downarrow)}{L\uparrow}{<-} + \RTDone{(\uparrow,\downarrow)}{L\uparrow}{R\downarrow}{(\uparrow,\downarrow)}{L\uparrow}{->}
	\\&=i\Gamma_N\left[\int \frac{f_R^-(\omega)e^{i\chi}}{\omega^+-(E_{(\uparrow,\downarrow)}-\varepsilon_{L})}+\int \frac{f_R^-(\omega)e^{i\chi}}{(E_{(\uparrow,\downarrow)}-\varepsilon_{L})-\omega^+} \right]\nonumber\\
	&=\Gamma_N f^-_{R}(E_{(\uparrow,\downarrow)}-\varepsilon_{L})e^{i\chi}
\end{align*}
\begin{align*}
	&W_{(\uparrow,\downarrow),L\uparrow}=	\RTDtwo{L\uparrow}{(\uparrow,\downarrow)}{R\downarrow}{L\uparrow}{(\uparrow,\downarrow)}{->} + \RTDone{L\uparrow}{(\uparrow,\downarrow)}{R\downarrow}{L\uparrow}{(\uparrow,\downarrow)}{<-} 
	\\&=i\Gamma_N\left[\int \frac{f_R^+(\omega)e^{-i\chi}}{(E_{(\uparrow,\downarrow)}-\varepsilon_{L})-\omega^+} +\int \frac{f_R^+(\omega)e^{-i\chi}}{\omega^+-(E_{(\uparrow,\downarrow)}-\varepsilon_{L})}\right]\nonumber\\
	&=\Gamma_N f^+_{R}(E_{(\uparrow,\downarrow)}-\varepsilon_{L})e^{-i\chi}
\end{align*}
\normalsize
\subsubsection*{3 -- Diagonal coherent rates $W^{A}_{B} \equiv W^{A,A}_{B,B}$}

The diagonal coherent rates account for energy-renormalization effects induced by virtual tunneling processes and couple the populations of states $\ket{A}$ and $\ket{B}$. They are given by
\begin{align*}
	W^{A}_{B}
	= - \frac{\Gamma_N}{2}
	\sum_{\eta,\sigma} \sum_{x} \sum_{p=\pm}
	\Big[
	\delta^{p}_{x,\eta\sigma}
	\, \varphi^{p}_{\eta}\!\left(p\big(E^{p}_{x}(\eta\sigma)-E_{\bar{x}}\big),\nu_x\right)
	\Big],
\end{align*}
where $x \in \{A,B\}$ and $\bar{x}$ denotes the complementary state ($\bar{A}=B$, $\bar{B}=A$). The quantity $E^{p}_{x}(\eta\sigma)$ is the energy of the many-body state obtained by \emph{adding} ($p=+1$) or \emph{removing} ($p=-1$) an electron with spin $\sigma$ in dot $\eta$ to or from the state $x$. The parameters $\nu_A=+1$ and $\nu_B=-1$, and the functions $\varphi^{\pm}_{\eta}(x,\xi)$ are defined in Eq.~\eqref{eq_varphi}. Finally, $\delta^{p}_{x,\eta\sigma}=1$ if the corresponding addition ($p=+1$) or removal ($p=-1$) process is allowed, and $\delta^{p}_{x,\eta\sigma}=0$ otherwise.

As an illustration, for $A=(\sigma,0)$ and $B=(\uparrow\downarrow,\sigma)$ one obtains
\small
\begin{align*}
	&W^{\sigma,0}_{\uparrow\downarrow,\sigma}
	= -\frac{\Gamma_N}{2} \Big[
	\varphi^{-}_{L}\big(E_{(\uparrow\downarrow,\sigma)}-E_{(\sigma,0)}^{-}(L\sigma), +1\big)
	\\&
	+\varphi^{+}_{L}\big(E_{(\sigma,0)}^{+}(L\bar\sigma)-E_{(\uparrow\downarrow,\sigma)}, +1\big)
	+\varphi^{+}_{R}\big(E_{(\sigma,0)}^{+}(R\uparrow)-E_{(\uparrow\downarrow,\sigma)}, +1\big)
	\\&
		+\varphi^{+}_{R}\big(E_{(\sigma,0)}^{+}(R\downarrow)-E_{(\uparrow\downarrow,\sigma)}, +1\big)
+\varphi^{-}_{L}\big(E_{(\sigma,0)}-E_{(\uparrow\downarrow,\sigma)}^{-}(L\uparrow), -1\big)
\\&
		+\varphi^{-}_{L}\big(E_{(\sigma,0)}-E_{(\uparrow\downarrow,\sigma)}^{-}(L\downarrow), -1\big)
	+
\varphi^{-}_{R}\big(E_{(\sigma,0)}-E_{(\uparrow\downarrow,\sigma)}^{-}(R\sigma), -1\big)
\\&
	+
	\varphi^{+}_{R}\big(E_{(\uparrow\downarrow,\sigma)}^{+}(R\bar\sigma)-E_{(\sigma,0)}, -1\big)
	\Big].
\end{align*}
\normalsize
\subsubsection*{4 -- Off-diagonal coherent rates $W^{A,A}_{B,D}$ and $W^{A,C}_{B,B}$}

Within a linear expansion in the tunneling and pairing couplings, the off-diagonal coherent rates for which either $A=C$ or $B=D$ acquire finite contributions. In this approximation, these rates take the form
\begin{align}
	W^{A,C}_{B,D}=-i\braket{A|H_{SC}|C}\delta_{B,D}+i\braket{D|H_{SC}|B}\delta_{A,C},
\end{align}
where $H_{SC}$ corresponding to the second and third terms of Eq.~(\ref{H_CPS}).

\subsubsection*{5 -- General off-diagonal coherent rates $W^{A,C}_{B,D}$ with $A \neq B \neq C \neq D$}

We now consider the fully off-diagonal coherent rates connecting four distinct many-body states. Such contributions arise only when the pairs of states $(A,C)$ and $(B,D)$ differ by a single electron, ensuring that the corresponding matrix elements of the creation and annihilation operators are finite. Two distinct cases must therefore be distinguished, depending on the relative particle numbers of the states involved.

\paragraph*{Case (i): $N(A)=N(C)+1$ and $N(B)=N(D)+1$}
\small
\begin{align}
	W^{A,C}_{B,D}&=     \RTDone{C}{A}{\etan}{D}{B}{<-}+	\RTDtwo{C}{A}{\etan}{D}{B}{->} \nonumber\\
	&=i\Gamma_N\bra{A}c^\dagger_{\etan}\ket{C}\bra{D}c_{\etan}\ket{B}\nonumber\\&\times\left[\int \frac{f^+(\omega)e^{-i\chi}}{\omega^+-(E_A-E_D)}+\int \frac{f^+(\omega)e^{-i\chi}}{(E_B-E_C)-\omega^+}\right]\nonumber\\
	&=\frac{\Gamma_N}{2}\bra{A}c^\dagger_{\etan}\ket{C}\bra{D}c_{\etan}\ket{B}\left(\varphi^{+}_{\eta}(E_A-E_D,+1)\right.
	\nonumber
	\\&\left.+(\varphi^{+}_{\eta}(E_B-E_C,-1)\right)e^{-i\chi}
\end{align} 
\normalsize
\paragraph*{Case (ii): $N(C)=N(A)+1$ and $N(D)=N(B)+1$}
\small
\begin{align}
	W^{A,C}_{B,D}&=    	\RTDtwo{C}{A}{\etan}{D}{B}{<-} + \RTDone{C}{A}{\etan}{D}{B}{->}\nonumber\\
	&=i\Gamma_N\bra{A}c_{\etan}\ket{C}\bra{D}c^\dagger_{\etan}\ket{B}\nonumber\\
	&\times\left[\int \frac{f^-(\omega)e^{i\chi}}{\omega^+-(E_C-E_B)}+\int \frac{f^-(\omega)e^{i\chi}}{(E_D-E_A)-\omega^+}\right]\nonumber\\
	&=\frac{\Gamma_N}{2}\bra{A}c_{\etan}\ket{C}\bra{D}c^\dagger_{\etan}\ket{B}\left(\varphi^{-}_{\eta}(E_C-E_B,-1)\right.\nonumber
	\\&\left.+(\varphi^{-}_{\eta}(E_D-E_A,+1)\right)e^{i\chi}
\end{align}
\normalsize
\subsection{Cooper pair splitter: Limit $U_{\alpha}\to \infty$.}

We now consider the limit $U_{\alpha}\to\infty$, corresponding to strong local Coulomb blockade in each quantum dot. In this regime, double occupancy of the individual dots is completely suppressed, which leads to a substantial reduction in the number of coherences contributing to transport. As a consequence, the structure of the rate matrix simplifies significantly and all nonvanishing rate elements can be evaluated explicitly in a computationally tractable manner. In order to simplify the notation throughout this subsection, we adopt the shorthand $\ket{\sigma\sigma'}\equiv\ket{\sigma,\sigma'}$, $\ket{L\sigma}\equiv\ket{\sigma,0}$, $\ket{R\sigma}\equiv\ket{0,\sigma}$, and $E\equiv E_{\sigma,\sigma'}=\varepsilon_L+\varepsilon_R+U_{LR}$ where $\sigma,\sigma'\in\{\uparrow,\downarrow\}$ are arbitrary spin indices. Since we compute the transition rates $W^{\xi_1\xi_1'}_{\xi_2\xi_2'}$ to first order in both $\Gamma_N$ and $\Gamma_C$, and at  $\delta = 0$, this implies that the rates have to be evaluated at $\varepsilon_{\eta\sigma}=-U_{LR}/2$, and $E_{\sigma,\sigma'}=0$. In the following we present the rates, the related diagrams and their analytical evaluation for all non-vanishing contributions.
\subsubsection*{Diagonal population rates $W_{A} \equiv W^{A,A}_{A,A}$}
\small
\begin{align}
&W_{0}=\sum_{\etan}\left[
\RTDthree{0}{\etan}{0}{\etan}{0}{0}{<-}+
\RTDfour{0}{0}{\etan}{0}{\etan}{0}{->}\right]
\nonumber\\
&= -i\Gamma_N\sum_{\etan}\left[   
\int \frac{f^+_{\eta}(\omega)}{\omega^+-(\varepsilon_{\eta})} +
\int \frac{f^+_{\eta}(\omega)}{(\varepsilon_{\eta})-\omega^+}\right]\nonumber\\ &
\times \bra{\etan}d^{\dagger}_{\etan}\ket{0}\bra{0}d_{\etan}\ket{\etan}=-2\Gamma_N\sum_\eta f^{+}_{\eta}(-U_{LR}/2).
\end{align}
\begin{align}
	&W_{\etan}=
	\RTDthree{\etan}{(\etan,\etab)}{\etan}{\etab}{\etan}{\etan}{<-}+
	\RTDthree{\etan}{(\etan,\bar\eta\sigma)}{\etan}{\bar\eta\sigma}{\etan}{\etan}{<-}\nonumber\\
	&\qquad\quad+
	\RTDfour{\etan}{\etan}{\etab}{\etan}{(\etan,\etab)}{\etan}{->}+
	\RTDfour{\etan}{\etan}{\bar\eta\sigma}{\etan}{(\etan,\bar\eta\sigma)}{\etan}{->}\nonumber\\
	&\qquad\quad+
	\RTDthree{\etan}{0}{\etan}{\etan}{\etan}{\etan}{->}+
	\RTDfour{\etan}{\etan}{\etan}{\etan}{0}{\etan}{<-}
	\nonumber\\
	&= -i\Gamma_N\left[   \left(
	\int \frac{f^+_{\bar\eta}(\omega)}{\omega^+-(E-\varepsilon_\eta)}+
		\int \frac{f^+_{\bar\eta}(\omega)}{(E-\varepsilon_\eta)-\omega^+}\right)\right.\nonumber\\
		&\qquad\times \bra{(\etan,\etab)}d^{\dagger}_{\etab}\ket{\etan}\bra{\etan}d_{\etab}\ket{(\etan,\etab)}\nonumber\\
	&\quad+\left(\int \frac{f^+_{\bar\eta}(\omega)}{\omega^+-(E-\varepsilon_\eta)}+
	\int d\omega\frac{f^+_{\bar\eta}(\omega)}{(E-\varepsilon_\eta)-\omega^+}
	\right)\nonumber\\
	&\qquad\times \bra{(\etan,\bar\eta\sigma)}d^{\dagger}_{\bar\eta\sigma}\ket{\etan}\bra{\etan}d_{\bar\eta\sigma}\ket{(\etan,\bar\eta\sigma)}\nonumber\\
	&\quad+\left(\int \frac{f^-_{\eta}(\omega)}{(\varepsilon_\eta)-\omega^+} 
	+\int \frac{f^-_{\eta}(\omega)}{\omega^+-(\varepsilon_\eta)}\right)\nonumber\\&\left.\qquad\times \bra{0}d_{\etan}\ket{\etan}\bra{\etan}d^{\dagger}_{\etan}\ket{0}\right]
	\nonumber\\
	&=-\left(W_{\ud,\etan}+W_{\du,\etan}+W_{\uu,\etan}+W_{\dd,\etan}+W_{0,\etan}\right)\big|_{\chi=0}\nonumber\\&=-2\Gamma_N f^{+}_{\bar\eta}(U_{LR}/2)-\Gamma_Nf^{-}_{\eta}(-U_{LR}/2).
\end{align}
\subsubsection*{Off-diagonal population rates $W_{A,B} \equiv W^{A,B}_{A,B}$}
\small
\begin{align}
	&W_{\etan,0} =	\RTDone{0}{\etan}{\etan}{0}{\etan}{<-} + 	\RTDtwo{0}{\etan}{\etan}{0}{\etan}{->}
	\nonumber	\\
	&=i\Gamma\left[\int \frac{f^+_{\eta}(\omega)e^{-i\chi}}{\omega^+-(\varepsilon_{\eta})} +\int \frac{f^+_{\eta}(\omega)e^{-i\chi}}{(\varepsilon_{\eta})-\omega^+}\right]\nonumber\\ &\times\bra{\etan}d^\dagger_{\etan}\ket{0}\bra{0}d_{\etan}\ket{\etan}=\Gamma_N f^+_{\eta}(-U_{LR}/2)e^{-i\chi}.
\end{align}
\begin{align}
	&W_{0,\etan} =	\RTDone{\etan}{0}{\etan}{\etan}{0}{->} + 	\RTDtwo{\etan}{0}{\etan}{\etan}{0}{<-}
	\nonumber	\\
	&=i\Gamma\left[\int \frac{f^-_{\eta}(\omega)e^{i\chi}}{(\varepsilon_{\eta})-\omega^+} +\int \frac{f^-_{\eta}(\omega)e^{i\chi}}{\omega^+-(\varepsilon_{\eta})}\right]\nonumber\\ &\times \bra{0}d_{\etan}\ket{\etan}\bra{\etan}d^\dagger_{\etan}\ket{0}	=\Gamma_N f^-_{\eta}(-U_{LR}/2)e^{i\chi}.
\end{align}
\begin{align}
	&W_{\etan,\ud} = 	\RTDtwo{\ud}{\etan}{\etab}{\ud}{\etan}{<-} + \RTDone{\ud}{\etan}{\etab}{\ud}{\etan}{->} 
	\nonumber	\\
	&=i\Gamma\left[\int \frac{f^-_{\bar\eta}(\omega)e^{i\chi}}{\omega^+-(E-\varepsilon_{\eta})} +\int \frac{f^-_{\bar\eta}(\omega)e^{i\chi}}{(E-\varepsilon_{\eta})-\omega^+}\right]\nonumber\\
	&\times \bra{\etan}d_{\etab}\ket{\ud}\bra{\ud}d^{\dagger}_{\etab}\ket{\etan}\nonumber\\
	&=\Gamma_N f^-_{\bar\eta}(U_{LR}/2)e^{i\chi}\left(\delta_{\etan,L\uparrow}+\delta_{\etan,R\downarrow}\right).
\end{align}
\begin{align}
		&W_{\ud,\etan}= 	\RTDtwo{\etan}{\ud}{\etab}{\etan}{\ud}{->} + \RTDone{\etan}{\ud}{\etab}{\etan}{\ud}{<-} 
	\nonumber	\\
	&=i\Gamma\left[\int \frac{f^+_{\bar\eta}(\omega)e^{-i\chi}}{(E-\varepsilon_{\eta})-\omega^+} +\int \frac{f^+(\omega)_{\bar\eta}e^{-i\chi}}{\omega^+-(E-\varepsilon_{\eta})}\right]\nonumber\\
	&\times \bra{\ud}d^{\dagger}_{\etab}\ket{\etan}\bra{\etan}d_{\etab}\ket{\ud}\nonumber\\
	&=\Gamma_N f^+_{\bar\eta}(U_{LR}/2)e^{-i\chi}\left(\delta_{\etan,L\uparrow}+\delta_{\etan,R\downarrow}\right).
\end{align}
\normalsize
Also, in an analogous manner
\begin{align*}
	W_{\etan,\du} &=\Gamma_N f^-_{\bar\eta}(U_{LR}/2)e^{i\chi}\left(\delta_{\etan,L\downarrow}+\delta_{\etan,R\uparrow}\right)\\
	W_{\du,\etan}&=\Gamma_N f^+_{\bar\eta}(U_{LR}/2)e^{-i\chi}\left(\delta_{\etan,L\downarrow}+\delta_{\etan,R\uparrow}\right)\\
		W_{\etan,\uu} &=\Gamma_N f^-_{\bar\eta}(U_{LR}/2)e^{i\chi}\left(\delta_{\etan,L\uparrow}+\delta_{\etan,R\uparrow}\right)\\
	W_{\uu,\etan}&=\Gamma_N f^+_{\bar\eta}(U_{LR}/2)e^{-i\chi}\left(\delta_{\etan,L\uparrow}+\delta_{\etan,R\uparrow}\right)\\
		W_{\etan,\dd} &=\Gamma_N f^-_{\bar\eta}(U_{LR}/2)e^{i\chi}\left(\delta_{\etan,L\downarrow}+\delta_{\etan,R\downarrow}\right)\\
	W_{\dd,\etan}&=\Gamma_N f^+_{\bar\eta}(U_{LR}/2)e^{-i\chi}\left(\delta_{\etan,L\downarrow}+\delta_{\etan,R\downarrow}\right).
\end{align*}
On the other hand we have
\small
\begin{align}
	&W_{\ud,\ud}=
	\RTDthree{\ud}{L\uparrow}{\ud}{R\downarrow}{\ud}{\ud}{->}+
	\RTDthree{\ud}{R\downarrow}{\ud}{L\uparrow}{\ud}{\ud}{->}\nonumber\\
	&\qquad\quad+\RTDfour{\ud}{\ud}{R\downarrow}{\ud}{L\uparrow}{\ud}{<-}+
	\RTDfour{\ud}{\ud}{L\uparrow}{\ud}{R\downarrow}{\ud}{<-}
	\nonumber\\
	&= -i\Gamma_N   \left(
	\int \frac{f^-_{R}(\omega)}{(E-\varepsilon_L)-\omega^+}+
	\int \frac{f^-_{R}(\omega)}{\omega^+-(E-\varepsilon_L)}\right)\nonumber\\
	&\qquad\times \bra{L\uparrow}d_{R\downarrow}\ket{\ud}\bra{\ud}d^{\dagger}_{R\downarrow}\ket{L\uparrow}\nonumber\\
	& -i\Gamma_N\left(\int \frac{f^-_{L}(\omega)}{(E-\varepsilon_R)-\omega^+}+
	\int \frac{f^-_{L}(\omega)}{\omega^+-(E-\varepsilon_R)}
	\right)\nonumber\\
	&\qquad \times \bra{R\downarrow}d_{L\uparrow}\ket{\ud}\bra{\ud}d^{\dagger}_{L\uparrow}\ket{R\downarrow}
	\nonumber\\
	&=-\Gamma_N \sum_{\eta} f^{-}_{\eta}(U_{LR}/2).
\end{align}
\normalsize
Analogously
\begin{align*}
	W_{\du,\du}=W_{\uu,\uu}=W_{\dd,\dd}=-\Gamma_N \sum_{\eta} f^{-}_{\eta}(U_{LR}/2).
\end{align*}
\subsubsection*{Coherent rates $W^{A,C}_{B,D}$}
The last rates to be computed are related to coherences, and read
\begin{align}
	&W^{0,\,\,0}_{\ud,\ud}
	=
		\RTDfour{0}{0}{R\downarrow}{\ud}{L\uparrow}{\ud}{<-}
		+
			\RTDfour{0}{0}{L\uparrow}{\ud}{R\downarrow}{\ud}{<-}\nonumber\\&\qquad\quad
			+
			\sum_{\etan}
	\RTDthree{0}{\etan}{0}{\etan}{\ud}{\ud}{<-}\nonumber\\
	&=-i\Gamma_N\left[   
	     \int \frac{f^-_{R}(\omega)}{\omega^++\varepsilon_{L}} \bra{L\uparrow}d_{R\downarrow}\ket{\ud}\bra{\ud}d^{\dagger}_{R\downarrow}\ket{L\uparrow}\right.\nonumber\\
	&+\int \frac{f^-_{L}(\omega)}{\omega^++\varepsilon_{R}} \bra{R\downarrow}d_{L\uparrow}\ket{\ud}\bra{\ud}d^{\dagger}_{L\uparrow}\ket{R\downarrow}\nonumber\\
	&\left.+
	2\sum_{\eta}\int \frac{f^+_{\eta}(\omega)}{\omega^+-(\varepsilon_{\eta}-E)} \bra{\etan}d^{\dagger}_{\etan}\ket{0}\bra{0}d_{\etan}\ket{\etan}\right]\nonumber\\
	&=-\frac{\Gamma_N}{2}\sum_{\eta}\left[ \varphi^-_{\eta}(U_{LR}/2,-1)+2\varphi^+_{\eta}(-U_{LR}/2,+1)\right]
\end{align}
\normalsize
And also  $W^{0,\,\,0}_{\du,\du}=W^{0,\,\,0}_{\ud,\ud}$, $W^{\ud,\ud}_{0,\,\,0}	=W^{\du,\du}_{0,\,\,0}=(W^{0,\,\,0}_{\ud,\ud})^*$.

\small
\begin{align}
	&W^{\ud, \ud}_{\du,\du}
	=
	\RTDthree{\ud}{L\uparrow}{\ud}{R\downarrow}{\du}{\du}{->}
	+
	\RTDthree{\ud}{R\downarrow}{\ud}{L\uparrow}{\du}{\du}{->}\nonumber\\
	&
	\qquad\quad+
	\RTDfour{\ud}{\ud}{R\uparrow}{\du}{L\downarrow}{\du}{<-}
	+
	\RTDfour{\ud}{\ud}{L\downarrow}{\du}{R\uparrow}{\du}{<-}\nonumber\\
	&=-i\Gamma_N\left[   
	\int \frac{f^-_{R}(\omega)}{(E-\varepsilon_L)-\omega^+} \bra{L\uparrow}d_{R\downarrow}\ket{\ud}\bra{\ud}d^{\dagger}_{R\downarrow}\ket{L\uparrow}\right.\nonumber\\&
	+\int \frac{f^-_{L}(\omega)}{(E-\varepsilon_R)-\omega^+} \bra{R\downarrow}d_{L\uparrow}\ket{\ud}\bra{\ud}d^{\dagger}_{L\uparrow}\ket{R\downarrow}\nonumber\\
	&+\int \frac{f^-_{R}(\omega)}{\omega^+-(E-\varepsilon_L)} \bra{L\downarrow}d_{R\uparrow}\ket{\ud}\bra{\ud}d^{\dagger}_{R\uparrow}\ket{L\downarrow}\nonumber\\&
	\left.+\int \frac{f^-_{L}(\omega)}{\omega^+-(E-\varepsilon_R)} \bra{R\uparrow}d_{L\downarrow}\ket{\ud}\bra{\ud}d^{\dagger}_{L\downarrow}\ket{R\uparrow}\right]\nonumber\\
	&=-\Gamma_N\sum_{\eta}f^{-}_{\eta}(U_{LR}/2)
\end{align}
\normalsize
and $	W^{\du,\du}_{\ud, \ud}=W^{\ud, \ud}_{\du,\du}$. Finally, the coupling to the superconducting lead generates pairing-induced coherent rates between states differing by two electrons. In the $U_{\alpha}\to\infty$ limit, these contributions are entirely determined by the effective pairing term and take the following simple form:
\small
	\begin{align}
	W^{0,0}_{0,\ud}&=W^{0,0}_{\ud,0}=W^{\ud,\ud}_{\ud,0}=W^{\ud,\ud}_{0,\ud}=
	W^{\du,0}_{0,0}=W^{\du,0}_{\du,\du}\nonumber\\&=W^{0,\du}_{0,0}=W^{0,\du}_{\du,\du}=
	W^{\du,0}_{\ud,\ud}=W^{\du,\du}_{\ud,0}=W^{0,\du}_{\ud,\ud}\nonumber\\&=W^{\du,\du}_{0,\ud}
	=i\frac{\Gamma_{S}}{2}\\
	W^{0,0}_{0,\du}&=W^{0,0}_{\du,0}=W^{\du,\du}_{\du,0}=W^{\du,\du}_{0,\du}=
	W^{\ud,0}_{0,0}=W^{\ud,0}_{\ud,\ud}\nonumber\\&=W^{0,\ud}_{0,0}=W^{0,\ud}_{\ud,\ud}=
	W^{\ud,\ud}_{\du,0}=W^{\ud,0}_{\du,\du}=W^{\ud,\ud}_{0,\du}\nonumber\\&=W^{0,\ud}_{\du,\du}
	=-i\frac{\Gamma_{S}}{2} \, .
\end{align}
\normalsize

\section{Green’s-function derivations: Hartree--Fock approximation}\label{App2}

To benchmark the real-time diagrammatic results in the regime of weak interactions and small bias, we compute the steady-state current, zero-frequency noise, and the corresponding TURs using a Green’s-function approach. The hierarchy of equations of motion is truncated at the Hartree--Fock (HF) level.
\label{Appendixtwo}
\subsection{Single Quantum Dot}

We start deriving the Green's function of the single quantum dot, with the Hamiltonian given in Eq.~(\ref{eq_QD}). Using the anticommutator relations $
\{d_\sigma,d_{\sigma'}^\dagger\}=\delta_{\sigma\sigma'},\quad
\{d_\sigma,d_{\sigma'}\}=\{d_\sigma^\dagger,d_{\sigma'}^\dagger\}=0
$
, we have the following identities
\begin{subequations}
	\begin{align}
	&[d_\sigma, d^\dagger_\sigma d_\sigma]=d_\sigma,\qquad
	[d_\sigma^\dagger, d_\sigma^\dagger d_\sigma] = -d_\sigma^\dagger,\\
	&[d_\sigma^\dagger, d_\uparrow d_\downarrow] = s_\sigma d_{\bar\sigma},\qquad
	[d_\sigma,d^\dagger_\uparrow d^\dagger_\downarrow] = s_\sigma d^\dagger_{\bar\sigma}
	\end{align}
	\label{eq_identities}
\end{subequations}
with $s_\uparrow=+1,\ s_\downarrow=-1.$

We define the \emph{retarded} Green's function in the time domain as
\begin{equation}
\langle\!\langle A;B\rangle\!\rangle^{r}(t)
\equiv -i\,\theta(t)\,\langle\{A(t),B(0)\}\rangle.
\end{equation}
Its Fourier transform is
\begin{equation}
\langle\!\langle A;B\rangle\!\rangle^{r}(\omega)
=\int_{-\infty}^{\infty} dt\, e^{i\omega t}\,\langle\!\langle A;B\rangle\!\rangle^{r}(t),
\end{equation}
which leads to the equation of motion (EOM)
\begin{align}
	\omega^+ \langle\langle A;B\rangle\rangle^r
	= \langle\{A,B\}\rangle
	+ \langle\langle [A,H];B\rangle\rangle^r
	\label{eq_EOM}
\end{align}
with $	\omega^+\equiv\omega+i0^+$ and we introduce the dot Nambu spinor $\Psi^\dagger=(d^\dagger_\uparrow,\ d_{\downarrow})$. In this basis,
\begin{align}
	G^r(\omega)& \equiv \langle\langle \Psi:\Psi^\dagger\rangle\rangle^r
	=
	\begin{pmatrix}
		\langle\langle d_\uparrow: d^\dagger_\uparrow\rangle\rangle^r &
		\langle\langle d_\uparrow: d_{\downarrow}\rangle\rangle^r \\
		\langle\langle d^\dagger_{\downarrow}: d^\dagger_{\uparrow}\rangle\rangle^r &
		\langle\langle d^\dagger_{\downarrow}: d_{\downarrow}\rangle\rangle^r
	\end{pmatrix}\nonumber\\&=
	\begin{pmatrix}
		G_{11} &
		F \\
		\bar F &
		G_{22}
	\end{pmatrix}
\end{align}

Taking into account \cref{eq_identities} one finds the commutators that appear in \cref{eq_EOM}
\begin{align*}
	[d_\sigma,H] &= \varepsilon d_\sigma + U d_\sigma n_{\bar\sigma}
	+ \frac{\Gamma_S}{2}(\delta_{\sigma\uparrow} d^\dagger_\downarrow - \delta_{\sigma\downarrow} d^\dagger_\uparrow)
	+ \sum_k t_N c_{k\sigma}, \\
	[d^\dagger_\sigma,H] &= -\varepsilon d^\dagger_\sigma - U n_{\bar\sigma} d^\dagger_\sigma
	+ \frac{\Gamma_S}{2}(\delta_{\sigma\uparrow} d_\downarrow - \delta_{\sigma\downarrow} d_\uparrow)
	+ \sum_k t_N c^\dagger_{k\sigma}.
\end{align*}
In the wide-band limit,  $	\Sigma_N^r(\omega) = \sum_k \frac{|t_N|^2}{\omega^+ - \epsilon_k}= -\frac{i}{2}\Gamma_N$, and $	\bar\Sigma_N^r(\omega)= \sum_k \frac{|t_N|^2}{\omega^+ + \epsilon_k} = -\frac{i}{2}\Gamma_N$, and being $\Gamma_N = 2\pi |t_N|^2 \rho_N $, one can derive the EOM for  $G^r(\omega)$ 
\begin{align*}
	\Big[(\omega^++\frac{i}{2}\Gamma_N)I_2-\varepsilon\tau_z-\frac{\Gamma_S}{2}\tau_x\Big]
	G^r(\omega)=
	I_2 +UK^r(\omega)
\end{align*}
where 
\begin{align}
	K^r(\omega)=
	\begin{pmatrix}
		\langle\langle d_\uparrow n_\downarrow: d_\uparrow^\dagger\rangle\rangle^r &
		\langle\langle d_\uparrow n_\downarrow: d_\downarrow\rangle\rangle^r\\
		-\langle\langle n_\uparrow d_\downarrow^\dagger: d_\uparrow^\dagger\rangle\rangle^r &
		-\langle\langle n_\uparrow d_\downarrow^\dagger: d_\downarrow\rangle\rangle^r
	\end{pmatrix}
\end{align}
and since  $ g^r(\omega)=\left[
\big(\omega^+ + \tfrac{i}{2}\Gamma_N\big)I_2
-\varepsilon\tau_z
-\tfrac{\Gamma_S}{2}\tau_x\right]^{-1}$ we have that 
\begin{align}
	\big[	g^r(\omega)\big]^{-1}
	G^r(\omega)&=
	I_2 +UK^r(\omega) \implies
	\nonumber\\ \big[ G^r(\omega)\big]^{-1}&=\big[g^r(\omega)\big]^{-1}-\Sigma_U^r(\omega)
\end{align}
where $	\Sigma_U^r(\omega)=U K^r(\omega)\big[G^r(\omega)\big]^{-1}$ is the many-body self-energy.

Within the HF approximation, the
Coulomb interaction is linearized by replacing the quartic operator
product by its mean-field contractions. The interaction term is approximated as
\[
\begin{aligned}
	U\,\hat n_\uparrow \hat n_\downarrow
	\;\approx\;&
	U\Big(
	\langle \hat n_\downarrow\rangle\,\hat n_\uparrow
	+\langle \hat n_\uparrow\rangle\,\hat n_\downarrow
	-\langle \hat d_\downarrow \hat d_\uparrow\rangle\,\hat d_\uparrow^\dagger \hat d_\downarrow^\dagger
	\nonumber\\&-\langle \hat d_\uparrow^\dagger \hat d_\downarrow^\dagger\rangle\,\hat d_\downarrow \hat d_\uparrow-
	\langle \hat n_\uparrow\rangle\langle \hat n_\downarrow\rangle
	+\big|\langle \hat d_\downarrow \hat d_\uparrow\rangle\big|^2
	\Big).
\end{aligned}
\]
This factorization neglects connected
three-body correlations and retains only pairwise contractions, effectively reducing the problem to a quadratic one with
renormalized parameters. 
For a general triple product of fermionic operators one
uses
$
\hat A \hat B \hat C
\;\approx\;
\hat A\,\langle \hat B \hat C\rangle
- \hat B\,\langle \hat A \hat C\rangle
+ \hat C\,\langle \hat A \hat B\rangle ,
$
where the minus signs follow from fermionic anticommutation relations.
Applying this to the correlated products entering the EOM yields
\begin{subequations}
	\begin{align*}
		\hat d_\uparrow \hat n_\downarrow
		= \hat d_\uparrow \hat d_\downarrow^\dagger \hat d_\downarrow
		&\;\approx\;
		\langle \hat n_\downarrow\rangle\, \hat d_\uparrow
		- \langle \hat d_\downarrow \hat d_\uparrow\rangle\, \hat d_\downarrow^\dagger ,
		\\
		\hat n_\uparrow \hat d_\downarrow^\dagger
		= \hat d_\uparrow^\dagger \hat d_\uparrow \hat d_\downarrow^\dagger
		&\;\approx\;
		-\,\langle \hat d_\uparrow^\dagger \hat d_\downarrow^\dagger\rangle\, \hat d_\uparrow
		+ \langle \hat n_\uparrow\rangle\, \hat d_\downarrow^\dagger ,
	\end{align*}
\end{subequations}

which after some algebra leads to the linear system  (we use $\langle \hat n_\uparrow\rangle =\langle \hat n_\downarrow\rangle =n/2$, 
and $\langle \hat d_\downarrow \hat d_\uparrow\rangle =\kappa$, 
$\langle \hat d_\uparrow^\dagger \hat d_\downarrow^\dagger\rangle =\kappa^*$)
\begin{align}
	&\big[G^r(\omega)\big]^{-1}
	= \big[g^r(\omega)\big]^{-1}
	- U\!\left(\tfrac{n}{2}\tau_z + \kappa\,\tau_+ + \kappa^{*}\tau_-\right)
	\nonumber\\[2pt]
	&=
	\Big(\omega + \tfrac{i}{2}\Gamma_N\Big)I_2
	-\tilde{\varepsilon}\,\tau_z
	-\mathrm{Re}\,\tilde{\Delta}\,\tau_x
	+\mathrm{Im}\,\tilde{\Delta}\,\tau_y,
	\label{eq_Gr_one}
\end{align}
where 
$
\tilde{\varepsilon} = \varepsilon + \tfrac{U n}{2}, 
$ and $  
\tilde{\Delta} = \tfrac{\Gamma_S}{2} - U\kappa,
$ and $\tau_\alpha$ are the Pauli matrices acting in Nambu space. 
From \cref{eq_Gr_one}, the Hartree–Fock self-energy reads
\begin{align}
	\Sigma_U^r(\omega)=
	U\begin{pmatrix}
		\langle n_\downarrow\rangle &
		\langle d_\downarrow d_\uparrow\rangle\\
		\langle d_\uparrow^\dagger d_\downarrow^\dagger\rangle &
		-\langle n_\uparrow\rangle
	\end{pmatrix}
	= U
	\begin{pmatrix}
		\tfrac{n}{2} & \kappa\\
		\kappa^{*} & -\tfrac{n}{2}
	\end{pmatrix}.
\end{align}

The inversion of \cref{eq_Gr_one} gives
\begin{align}
	G^r(\omega)
	= \frac{1}{D}
	\begin{pmatrix}
		\omega+\tfrac{i}{2}\Gamma_N+\tilde{\varepsilon} & -\tilde{\Delta}\\[3pt]
		-\tilde{\Delta}^{*} & \omega+\tfrac{i}{2}\Gamma_N-\tilde{\varepsilon}
	\end{pmatrix},
	\label{eq_Gr_final}
\end{align}
where $D=(\omega+\tfrac{i}{2}\Gamma_N)^2 -E^2$ and  $E=\sqrt{\tilde{\varepsilon}^{\,2}+|\tilde{\Delta}|^{2}}$.
In the non-interacting limit \(U\to0\), one recovers 
\(\tilde{\varepsilon}=\varepsilon\) and \(\tilde{\Delta}=\tfrac{\Gamma_S}{2}\),
as expected.
\paragraph*{Integral of $n$ and $\kappa$:}
 In the steady–state regime, and within the HF approximation where $\Sigma_U^{<}=0$, the lesser Green’s function is $
	G^{<}(\omega)=G^{r}(\omega)\,\Sigma_N^{<}(\omega)\,G^{a}(\omega),
$
with the normal-lead lesser self–energy (Nambu basis $\Psi=(d_\uparrow,\,d_\downarrow^\dagger)^T$)
\begin{equation*}
	\Sigma_N^{<}(\omega)= i\,\Gamma_N
	\begin{pmatrix}
		f_e & 0\\
		0 & f_h
	\end{pmatrix},
	\qquad
	f_{e/h}=\frac{1}{e^{(\omega\mp\mu_N)/T}+1}.
\end{equation*}

Using $G^{a}=(G^{r})^\dagger$ one obtains
\begin{widetext}
\begin{align}
	G^{<}(\omega)
	= i\,\Gamma_N
	\begin{pmatrix}
		f_e\,|G_{11}^{r}|^{2}+f_h\,|G_{12}^{r}|^{2}
		& f_e\,G_{11}^{r}(G_{21}^{r})^{*}+f_h\,G_{12}^{r}(G_{22}^{r})^{*}\\[6pt]
		f_e\,G_{21}^{r}(G_{11}^{r})^{*}+f_h\,G_{22}^{r}(G_{12}^{r})^{*}
		& f_e\,|G_{21}^{r}|^{2}+f_h\,|G_{22}^{r}|^{2}
	\end{pmatrix}.
	\label{eq:Gless-matrix}
\end{align}
\end{widetext}
The density and the pairing form a system of integral coupled equations
\begin{subequations}
	\begin{align*}
		n
		&= -\,\frac{i}{\pi}\int d\omega\, G^{<}_{11}(\omega)
			\nonumber\\&= \frac{\Gamma_N}{\pi}\int d\omega\;
		\Big[
		f_e(\omega)\,|G_{11}^{r}(\omega)|^{2}
	+ f_h(\omega)\,|G_{12}^{r}(\omega)|^{2}
		\Big]
		\\
		\kappa
		&= -\,\frac{i}{2\pi}\int d\omega\, G^{<}_{12}(\omega)
			\nonumber\\&= \frac{\Gamma_N}{2\pi}\int d\omega\;
		\Big[
		f_e(\omega)\,G_{11}^{r}(\omega)\big(G_{21}^{r}(\omega)\big)^{*}
			\nonumber\\&\qquad+ f_h(\omega)\,G_{12}^{r}(\omega)\big(G_{22}^{r}(\omega)\big)^{*}
		\Big]
	\end{align*}
\end{subequations}
since $G^r$ depends on $n$ and $\kappa$ through $\tilde\varepsilon$ and $\tilde\Delta$. In order to analytically solve these integrals we express each of the integrands in the form of Lorentzian times Fermi or Lorentzian times difference of Fermis times $\omega$. In order to simplify the notation we define
\begin{align*}
	&A_\pm=\frac{1}{2}\Big(1\pm \frac{\tilde\varepsilon}{E}\Big),\qquad
	L_{\pm}(\omega)=\frac{1}{(\omega\mp E)^2+\tfrac{\Gamma_N^2}{4}},\\ &\xi = \frac{1}{4\left(E^2+\tfrac{\Gamma_N^2}{4}\right)},\qquad
	E=\sqrt{\tilde\varepsilon^{\,2}+\tilde\Delta^{\,2}},
\end{align*}

\noindent
Using the two useful identities  (we ommit the $\omega$ dependence for brevity)
\begin{align*}
	L_+L_-
	&=\xi
	\left[
	2\big(L_++L_-\big)
	-\frac{\omega}{E}\big(L_+-L_-\big)
	\right],\\
	\frac{\omega^{2}}{E}\big(L_+-L_-\big)
	&= 2\,\omega\,\big(L_+ + L_-\big)\;-\;\frac{E^2+\Gamma_N^2/4}{E}\,\big(L_+-L_-\big),
\end{align*}
the relevant quantities read
\begin{subequations}
	\begin{align*}
		|G_{11}^r|^2
		&= \sum_{s=\pm1}
		\Big[
		\big(A_s - 2\xi\,\tilde\Delta^{\,2}\big)
		+ s\,\frac{\xi\,\tilde\Delta^{\,2}}{E}\,\omega
		\Big]\,L_s(\omega),
		\\
		|G_{12}^r|^2
		&= \sum_{s=\pm1}
		\Big[
		2\xi\,\tilde\Delta^{\,2}
		- s\,\frac{\xi\,\tilde\Delta^{\,2}}{E}\,\omega
		\Big]\,L_s(\omega).
		\\
		G_{11}^r(G_{21}^r)^{*}
		&= -\,\tilde\Delta\,\xi
		\sum_{s=\pm1}
		\Big[
		2(\tilde\varepsilon+i\Gamma_N/2)
		\nonumber\\&+ s\,\frac{E^{2}+\Gamma_N^2/4}{E}
		- s\,\frac{(\tilde\varepsilon+i\Gamma_N/2)}{E}\,\omega
		\Big]\,L_s(\omega).
		\\
		G_{12}^r(G_{22}^r)^{*}
		&= -\,\tilde\Delta\,\xi
		\sum_{s=\pm1}
		\Big[
		-2(\tilde\varepsilon+i\Gamma_N/2)
		\nonumber\\&+ s\,\frac{E^{2}+\Gamma_N^2/4}{E}
		+ s\,\frac{(\tilde\varepsilon+i\Gamma_N/2)}{E}\,\omega
		\Big]\,L_s(\omega).
	\end{align*}
\end{subequations}

Using the auxiliary integrals \cite{sobrino2021thermoelectric}
\begin{align*}
	\mathcal I_{0}^{\sigma,s}
	&=\int
	\frac{f_\sigma(\omega)}{(\omega-sE)^2+\Gamma_N^2/4}\,d\omega
	= \frac{\pi - 2\,\text{Im}\left[\psi(z_{\sigma,s})\right]}{\Gamma_N},\\
	\mathcal J_{0}^{eh,s}
	&=\int
	\frac{\omega\,[f_e(\omega)-f_h(\omega)]}{(\omega-sE)^2+\Gamma_N^2/4}\,d\omega
	= \text{Re}\left\{\psi(z_{e,s})-\psi(z_{h,s})\right\}
	\nonumber\\&\;+\; sE\,\big(\mathcal I_{0}^{e,s}-\mathcal I_{0}^{h,s}\big),
\end{align*}

where \(\psi(z)\) is the digamma function, $z_{\sigma,s}=\tfrac{1}{2}+\tfrac{\Gamma_N/2+i(sE-\mu_\sigma)}{2\pi T}$, and \(\mu_e=\mu_N\), \(\mu_h=-\mu_N\), we find 
\begin{widetext}
\begin{subequations}
	\begin{align}
		n
		&=
		\frac{\Gamma_N}{\pi}
		\sum_{s=\pm1}
		\left[
		\big(A_s - 2\,\xi\,\tilde{\Delta}^{\,2}\big)\,
		\mathcal{I}_{0}^{e,s}
		+ 2\,\xi\,\tilde{\Delta}^{\,2}\,
		\mathcal{I}_{0}^{h,s}
		+ \frac{s\,\xi\,\tilde{\Delta}^{\,2}}{E}\,
		\mathcal{J}_{0}^{eh,s}
		\right],
		\\[10pt]
		\kappa
		&=
		-\,\frac{\Gamma_N\,\tilde{\Delta}\,\xi}{2\pi}
		\sum_{s=\pm1}
		\left[
		\Big(2\,\tilde{\varepsilon} + s\,\frac{E^{2}+\gamma^{2}}{E}\Big)\,
		\mathcal{I}_{0}^{e,s}
		+\Big(-2\,\tilde{\varepsilon} + s\,\frac{E^{2}+\gamma^{2}}{E}\Big)\,
		\mathcal{I}_{0}^{h,s}
		-\frac{s\,\tilde{\varepsilon}}{E}\,
		\mathcal{J}_{0}^{eh,s}
		\right].
	\end{align}
	\label{n_k_integrated}
\end{subequations}
Eqs.~(\ref{n_k_integrated}) form a system of coupled equations that need to be solved self-consistently. The current and the noise can be directly computed through \cite{mayo2025thermodynamic}
\begin{subequations}
\begin{align*}
	I 
	&= -\,\frac{\Gamma_N}{2\pi} 
	\int_{-\infty}^{\infty} d\omega\;
	\text{Im}\Big\{
	\operatorname{Tr}\left[
	\tau_3 \big( 2 F^+(\omega) G^r(\omega) + G^<(\omega) \big)
	\right]
	\Big\} = \frac{\Gamma_N^2}{\pi} 
	\int_{-\infty}^{\infty} d\omega\;
	\,\big|G^r_{12}(\omega)\big|^2\,
	\Big[ f(\omega - \mu_N) - f(\omega + \mu_N) \Big] ,
\\
	S(0)
	&= \frac{1}{2\pi}
	\int_{-\infty}^{+\infty}  d\omega\;
	\operatorname{Re}\Bigg\{
	\Gamma_N\,\mathrm{Tr}\left[
	i F_+(\omega) G^{>}(\omega)
	- i F_-(\omega) G^{<}(\omega)
	\right]
	+\frac{\Gamma_N^2}{4}\,
	\mathrm{Tr}\Big[
	\tau_3 (2 G^r F_+ - 2 F_+ G^a + G^{<}) \tau_3 G^{>}
	\nonumber\\[3pt]
	&\hspace{2.8cm}
	-\,\tau_3 (2 G^r F_- - 2 F_- G^a - G^{>}) \tau_3 G^{<}
	- 2\,\tau_3 (2 G^r F_+ + G^{<}) \tau_3 (2 G^r F_- - G^{>})
	\Big]
	\Bigg\}
\end{align*}
\end{subequations}
\end{widetext}

\subsection{Cooper Pair Splitter}
For the CPS system we proceed analogously to the single quantum dot case. Using the canonical anticommutation relations
$\{d_{\alpha\sigma},d^\dagger_{\beta\sigma'}\}=\delta_{\alpha\beta}\delta_{\sigma\sigma'}$ and
$\{d_{\alpha\sigma},d_{\beta\sigma'}\}=0$, together with the Hamiltonian in Eq.~(\ref{H_CPS}), one obtains the following commutation identities:
\begin{align*}
	[d_{\alpha\sigma},\, d^\dagger_{\alpha\sigma} d_{\alpha\sigma}]
	&= d_{\alpha\sigma},
	&
	[d^\dagger_{\alpha\sigma},\, d^\dagger_{\alpha\sigma} d_{\alpha\sigma}]
	&= -\,d^\dagger_{\alpha\sigma},  \\
	[d^\dagger_{\alpha\sigma},\, d_{\alpha\uparrow} d_{\alpha\downarrow}]
	&= s_\sigma\, d_{\alpha\bar\sigma},
	&
	[d_{\alpha\sigma},\, d^\dagger_{\alpha\uparrow} d^\dagger_{\alpha\downarrow}]
	&= s_\sigma\, d^\dagger_{\alpha\bar\sigma},  \\
	[d_{\eta\sigma},\, H]
	&= s_\eta s_\sigma \, d^\dagger_{\bar\eta\bar\sigma},
	&
	[d^\dagger_{\eta\sigma},\, H]
	&= -\,s_\eta s_\sigma \, d_{\bar\eta\bar\sigma}.
\end{align*}
Here $s_\uparrow=+1$, $s_\downarrow=-1$, $s_L=+1$, $s_R=-1$, and $\bar\eta$ ($\bar\sigma$) denotes the opposite lead (spin).
The Nambu spinor in the CPS is $	\Psi^\dagger \equiv \big(d^\dagger_{L\uparrow},\, d_{L\downarrow},\, d^\dagger_{R\uparrow},\, d_{R\downarrow}\big)$, 
and the GF in the Nambu dot basis reads
\begin{widetext}
\begin{equation*}
	G^r(\omega)\equiv\langle\langle \Psi:\Psi^\dagger\rangle\rangle^r_\omega=
	\begin{pmatrix}
		G^{LL}_{11} & F^{LL} & G^{LR}_{11} & F^{LR} \\
		\bar F^{LL} & G^{LL}_{22} & \bar F^{LR} & G^{LR}_{22} \\
		G^{RL}_{11} & F^{RL} & G^{RR}_{11} & F^{RR} \\
		\bar F^{RL} & G^{RL}_{22} & \bar F^{RR} & G^{RR}_{22}
	\end{pmatrix}=
	\begin{pmatrix}
		\langle\langle d_{L\uparrow} :d^\dagger_{L\uparrow}\rangle\rangle^r
		&
		\langle\langle d_{L\uparrow}:d_{L\downarrow}\rangle\rangle^r
		&
		\langle\langle d_{L\uparrow}:d^\dagger_{R\uparrow}\rangle\rangle^r
		&
		\langle\langle d_{L\uparrow}:d_{R\downarrow}\rangle\rangle^r
		\\
		\langle\langle d^\dagger_{L\downarrow}:d^\dagger_{L\uparrow}\rangle\rangle^r
		&
		\langle\langle d^\dagger_{L\downarrow}:d_{L\downarrow}\rangle\rangle^r
		&
		\langle\langle d^\dagger_{L\downarrow}:d^\dagger_{R\uparrow}\rangle\rangle^r
		&
		\langle\langle d^\dagger_{L\downarrow}:d_{R\downarrow}\rangle\rangle^r
		\\
		\langle\langle d_{R\uparrow}:d^\dagger_{L\uparrow}\rangle\rangle^r
		&
		\langle\langle d_{R\uparrow}:d_{L\downarrow}\rangle\rangle^r
		&
		\langle\langle d_{R\uparrow}:d^\dagger_{R\uparrow}\rangle\rangle^r
		&
		\langle\langle d_{R\uparrow}:d_{R\downarrow}\rangle\rangle^r
		\\
		\langle\langle d^\dagger_{R\downarrow}:d^\dagger_{L\uparrow}\rangle\rangle^r
		&
		\langle\langle d^\dagger_{R\downarrow}:d_{L\downarrow}\rangle\rangle^r
		&
		\langle\langle d^\dagger_{R\downarrow}:d^\dagger_{R\uparrow}\rangle\rangle^r
		&
		\langle\langle d^\dagger_{R\downarrow}:d_{R\downarrow}\rangle\rangle^r
	\end{pmatrix}.
\end{equation*}
\end{widetext}
The EOM reads 
\begin{align*}
	(g^r)^{-1} G^r(\omega)
	=
	\mathbb{I}
	+
	U_L\,K^{L}(\omega)
	+
	U_R\,K^{R}(\omega)
	+
	U_{LR}\,K^{LR}(\omega)
\end{align*}
\noindent
\begin{figure*}
	\centering
	\includegraphics[width=0.9\linewidth]{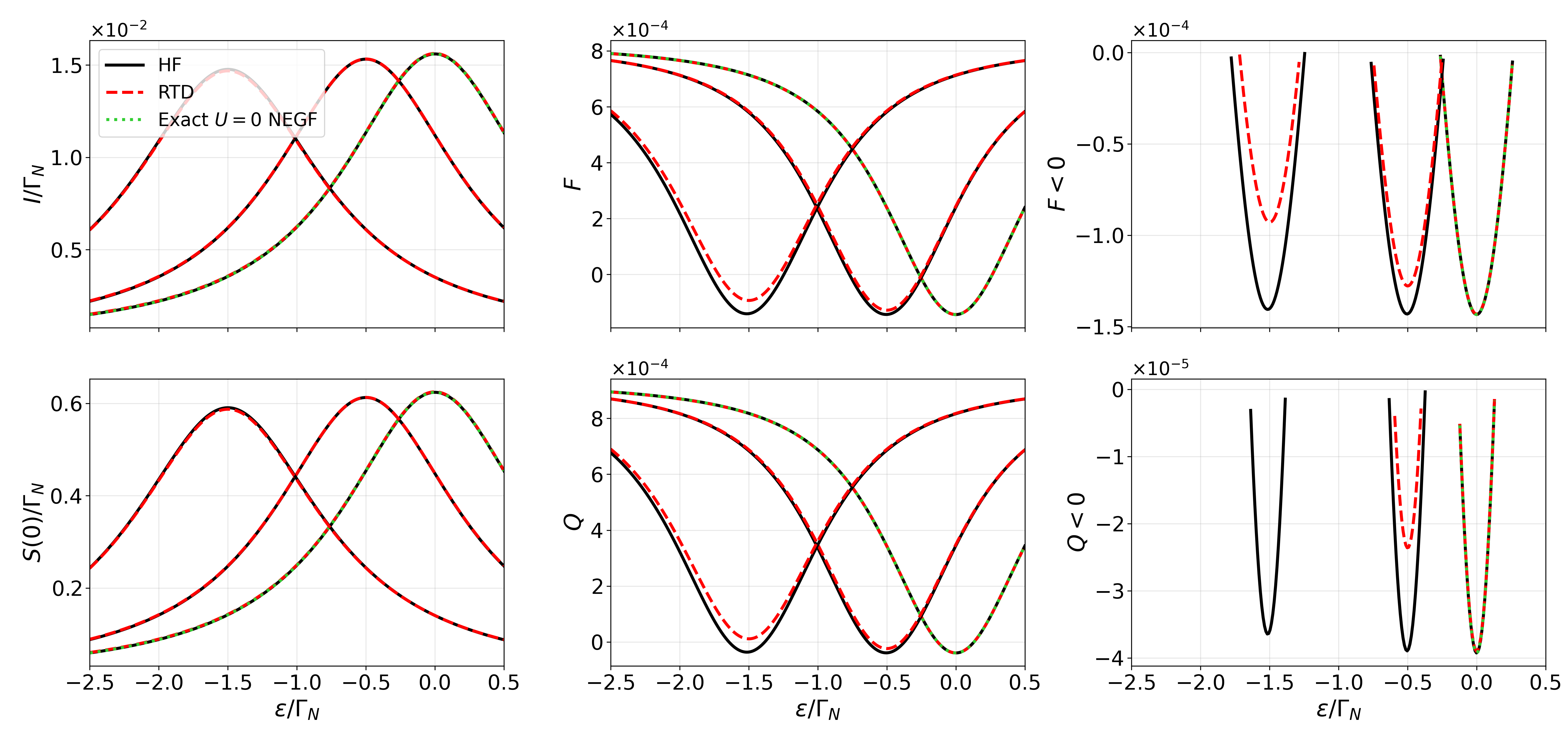}
	\caption{
		Superconducting current, zero-frequency noise, and classical and quantumTURs of a single quantum dot, shown as functions of the level position $\varepsilon/(k_B T)$ for $\mu_N = 0.5\,\Gamma_N$, $\Gamma_S = \sqrt{5/3}\,\Gamma_N$, and $k_B T =10\Gamma_N$. Curves correspond to Coulomb interactions $U/\Gamma_N = 0, 1, 3$, with increasing $U$ shifting the features toward lower $\varepsilon$. 
		Left column: current $I$ (top) and noise $S(0)$ (bottom). 
		Middle column: classical and quantum TURs, $F$ and $Q$. 
		Right column: negative contributions of $F$ and $Q$, highlighting violation regimes. 
		Solid black lines denote the Hartree--Fock approximation, dashed red lines the real-time diagrammatic approach, and dotted green lines the exact nonequilibrium Green's-function result, available only for $U=0$. 
		All energies in units of  $\Gamma_N$.
	}
	\label{figure11}
\end{figure*}
or equivalently
\begin{align*}
	G^r(\omega)=
	g^r(\omega)+g^r(\omega)\,\Sigma_{\mathrm{int}}^r(\omega)\,G^r(\omega)
\end{align*}

where we have used  $
\Sigma_{\mathrm{int}}^r(\omega)\,G^r(\omega)
\;\equiv\;
U_L\,K^{L}(\omega)
\;+\;
U_R\,K^{R}(\omega)
\;+\;
U_{LR}\,K^{LR}(\omega)$
, with the non-interacting GF given by 
\begin{equation}
	g^r(\omega)
	=
	\left[
	\begin{pmatrix}
		\tilde\omega - \varepsilon_L & \frac{\Gamma_S}{2} & 0 & \frac{\Gamma_C}{2} \\
		\frac{\Gamma_S}{2} & \tilde\omega + \varepsilon_L & \frac{\Gamma_C}{2} & 0 \\
		0 & \frac{\Gamma_C}{2} & \tilde\omega - \varepsilon_R & \frac{\Gamma_S}{2} \\
		\frac{\Gamma_C}{2} & 0 & \frac{\Gamma_S}{2} & \tilde\omega + \varepsilon_R
	\end{pmatrix}
	\right]^{-1},
	\label{eq:g_inv_WBL}
\end{equation}
where $\tilde\omega = \omega + i\Gamma_N/2$, and $K^{(L/R/LR)}(\omega)$ collecting the higher–order GF generated by the interaction terms
\begin{widetext}v
\begin{subequations}
	\begin{equation*}
		K^{L}(\omega)=
		\begin{pmatrix}
			\langle\langle n_{L\downarrow} d_{L\uparrow} : d^\dagger_{L\uparrow}\rangle\rangle	&
			\langle\langle n_{L\downarrow} d_{L\uparrow} : d_{L\downarrow}\rangle\rangle&
			\langle\langle n_{L\downarrow} d_{L\uparrow} : d^\dagger_{R\uparrow}\rangle\rangle&
			\langle\langle n_{L\downarrow} d_{L\uparrow} : d_{R\downarrow}\rangle\rangle	\\
			-\,\langle\langle n_{L\uparrow} d^\dagger_{L\downarrow} : d^\dagger_{L\uparrow}\rangle\rangle&
			-\,\langle\langle n_{L\uparrow} d^\dagger_{L\downarrow} : d_{L\downarrow}\rangle\rangle&
			-\,\langle\langle n_{L\uparrow} d^\dagger_{L\downarrow} : d^\dagger_{R\uparrow}\rangle\rangle&
			-\,\langle\langle n_{L\uparrow} d^\dagger_{L\downarrow} : d_{R\downarrow}\rangle\rangle
			\\
			0&0&0&0\\
			0&0&0&0
		\end{pmatrix}.
	\end{equation*}
	\begin{equation*}
		K^{R}(\omega)=
		\begin{pmatrix}
			0&0&0&0\\
			0&0&0&0\\
			\langle\langle n_{R\downarrow} d_{R\uparrow} : d^\dagger_{L\uparrow}\rangle\rangle&
			\langle\langle n_{R\downarrow} d_{R\uparrow} : d_{L\downarrow}\rangle\rangle&
			\langle\langle n_{R\downarrow} d_{R\uparrow} : d^\dagger_{R\uparrow}\rangle\rangle&
			\langle\langle n_{R\downarrow} d_{R\uparrow} : d_{R\downarrow}\rangle\rangle\\
			-\,\langle\langle n_{R\uparrow} d^\dagger_{R\downarrow} : d^\dagger_{L\uparrow}\rangle\rangle&
			-\,\langle\langle n_{R\uparrow} d^\dagger_{R\downarrow} : d_{L\downarrow}\rangle\rangle&
			-\,\langle\langle n_{R\uparrow} d^\dagger_{R\downarrow} : d^\dagger_{R\uparrow}\rangle\rangle&
			-\,\langle\langle n_{R\uparrow} d^\dagger_{R\downarrow} : d_{R\downarrow}\rangle\rangle
		\end{pmatrix}
	\end{equation*}
	\begin{equation*}
		K^{LR}(\omega)=
		\begin{pmatrix}
			\langle\langle n_{R}\, d_{L\uparrow} : d^\dagger_{L\uparrow}\rangle\rangle &
			\langle\langle n_{R}\, d_{L\uparrow :} d_{L\downarrow}\rangle\rangle &
			\langle\langle n_{R}\, d_{L\uparrow} : d^\dagger_{R\uparrow}\rangle\rangle &
			\langle\langle n_{R}\, d_{L\uparrow} : d_{R\downarrow}\rangle\rangle	\\[0.6ex]
			-\,\langle\langle n_{R}\, d^\dagger_{L\downarrow} : d^\dagger_{L\uparrow}\rangle\rangle &
			-\,\langle\langle n_{R}\, d^\dagger_{L\downarrow} : d_{L\downarrow}\rangle\rangle &
			-\,\langle\langle n_{R}\, d^\dagger_{L\downarrow} : d^\dagger_{R\uparrow}\rangle\rangle &
			-\,\langle\langle n_{R}\, d^\dagger_{L\downarrow} : d_{R\downarrow}\rangle\rangle\\[0.6ex]
			\langle\langle n_{L}\, d_{R\uparrow} : d^\dagger_{L\uparrow}\rangle\rangle &
			\langle\langle n_{L}\, d_{R\uparrow} : d_{L\downarrow}\rangle\rangle &
			\langle\langle n_{L}\, d_{R\uparrow} : d^\dagger_{R\uparrow}\rangle\rangle &
			\langle\langle n_{L}\, d_{R\uparrow} : d_{R\downarrow}\rangle\rangle\\[0.6ex]
			-\,\langle\langle n_{L}\, d^\dagger_{R\downarrow} : d^\dagger_{L\uparrow}\rangle\rangle &
			-\,\langle\langle n_{L}\, d^\dagger_{R\downarrow} : d_{L\downarrow}\rangle\rangle &
			-\,\langle\langle n_{L}\, d^\dagger_{R\downarrow} : d^\dagger_{R\uparrow}\rangle\rangle &
			-\,\langle\langle n_{L}\, d^\dagger_{R\downarrow} : d_{R\downarrow}\rangle\rangle
		\end{pmatrix}.
	\end{equation*}
\end{subequations}
\end{widetext}
\begin{figure*}
	\includegraphics[width=0.9\linewidth]{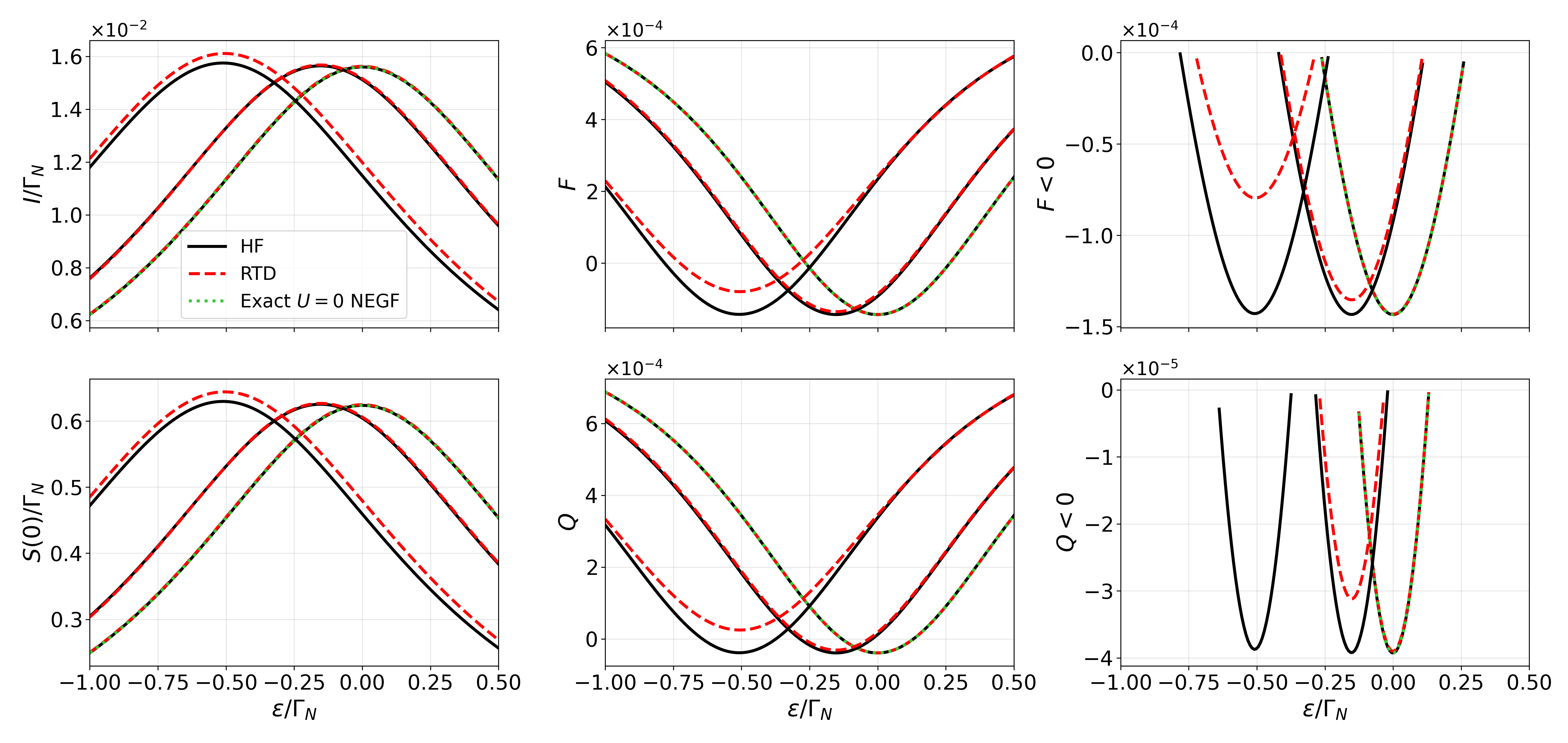}
	\caption{Superconducting current, zero-frequency noise, and classical and quantum TURs of a Cooper-pair splitter, shown as functions of the common gate level $\varepsilon/\Gamma_N$ for $\mu_N = 0.5\,\Gamma_N$, $\Gamma_S = \Gamma_C = \sqrt{5/3}\,\Gamma_N/2$, and $k_B T =10\Gamma_N$. Curves correspond to Coulomb interactions $U/\Gamma_N = 0, 0.3, 1$. 
		Solid black lines denote the Hartree--Fock approximation, dashed red lines the real-time diagrammatic approach, and dotted green lines the exact nonequilibrium Green’s-function result, available only for $U=0$. 
		All energies in units of  $\Gamma_N$.}
	\label{figure12}
\end{figure*}
Within the HF approximation, assuming spin symmetry,
$\langle n_{\alpha\uparrow}\rangle=\langle n_{\alpha\downarrow}\rangle\equiv n_\alpha/2$,
and introducing local and nonlocal pairing amplitudes
$\kappa_\alpha\equiv\langle d_{\alpha\downarrow}d_{\alpha\uparrow}\rangle$ ($\alpha=L,R$),
together with $\kappa_{LR}=\langle d_{R\downarrow}d_{L\uparrow}\rangle$, and $
\kappa_{RL}=\langle d_{L\downarrow}d_{R\uparrow}\rangle$ for interdot pairing correlations, the correlated GFs appearing in the EOM are approximated by their HF decouplings. For instance one finds $\big\langle\!\big\langle d_{L\uparrow}\, n_{L\downarrow} : B \big\rangle\!\big\rangle
\approx
\frac{n_L}{2}\,\big\langle\!\big\langle d_{L\uparrow} : B \big\rangle\!\big\rangle
\;+\;
\kappa_{L}\,\big\langle\!\big\langle d^\dagger_{L\downarrow} : B \big\rangle\!\big\rangle$. 
Substituting in the $K$ matrices one obtains 
\begin{widetext}
\begin{subequations}
	\begin{equation*}
		K^{L}_{\mathrm{HF}}(\omega)=
		\begin{pmatrix}
			\frac{n_L}{2}\,G^{LL}_{11} + \kappa_{L}\,\bar F^{LL}
			&
			\frac{n_L}{2}\,F^{LL} + \kappa_{L}\,G^{LL}_{22}
			&
			\frac{n_L}{2}\,G^{LR}_{11} + \kappa_{L}\,\bar F^{LR}
			&
			\frac{n_L}{2}\,F^{LR} + \kappa_{L}\,G^{LR}_{22}
			\\[0.8ex]
			-\,\frac{n_L}{2}\,\bar F^{LL} + \kappa_{L}^{*}\,G^{LL}_{11}
			&
			-\,\frac{n_L}{2}\,G^{LL}_{22} + \kappa_{L}^{*}\,F^{LL}
			&
			-\,\frac{n_L}{2}\,\bar F^{LR} + \kappa_{L}^{*}\,G^{LR}_{11}
			&
			-\,\frac{n_L}{2}\,G^{LR}_{22} + \kappa_{L}^{*}\,F^{LR}
			\\[0.8ex]
			0 & 0 & 0 & 0\\
			0 & 0 & 0 & 0
		\end{pmatrix},
	\end{equation*}
	\begin{equation*}
		K^{R}_{\mathrm{HF}}(\omega)=
		\begin{pmatrix}
			0 & 0 & 0 & 0\\
			0 & 0 & 0 & 0\\[0.8ex]
			\frac{n_R}{2}\,G^{RL}_{11} + \kappa_{R}\,\bar F^{RL}
			&
			\frac{n_R}{2}\,F^{RL} + \kappa_{R}\,G^{RL}_{22}
			&
			\frac{n_R}{2}\,G^{RR}_{11} + \kappa_{R}\,\bar F^{RR}
			&
			\frac{n_R}{2}\,F^{RR} + \kappa_{R}\,G^{RR}_{22}
			\\[0.8ex]
			-\,\frac{n_R}{2}\,\bar F^{RL} + \kappa_{R}^{*}\,G^{RL}_{11}
			&
			-\,\frac{n_R}{2}\,G^{RL}_{22} + \kappa_{R}^{*}\,F^{RL}
			&
			-\,\frac{n_R}{2}\,\bar F^{RR} + \kappa_{R}^{*}\,G^{RR}_{11}
			&
			-\,\frac{n_R}{2}\,G^{RR}_{22} + \kappa_{R}^{*}\,F^{RR}
		\end{pmatrix},
	\end{equation*}
	\begin{equation*}
		K^{LR}_{\mathrm{HF}}(\omega)=
		\begin{pmatrix}
			n_{R}\,G^{LL}_{11} + \kappa_{LR}\,\bar F^{RR}
			&
			n_{R}\,F^{LL} + \kappa_{LR}\,G^{RR}_{22}
			&
			n_{R}\,G^{LR}_{11} + \kappa_{LR}\,\bar F^{LR}
			&
			n_{R}\,F^{LR} + \kappa_{LR}\,G^{LR}_{22}
			\\[0.8ex]
			-\,n_{R}\,\bar F^{LL} + \kappa_{RL}^{*}\,G^{RR}_{11}
			&
			-\,n_{R}\,G^{LL}_{22} + \kappa_{RL}^{*}\,F^{RR}
			&
			-\,n_{R}\,\bar F^{LR} + \kappa_{RL}^{*}\,G^{RL}_{11}
			&
			-\,n_{R}\,G^{LR}_{22} + \kappa_{RL}^{*}\,F^{RL}
			\\[0.8ex]
			n_{L}\,G^{RL}_{11} + \kappa_{RL}\,\bar F^{LL}
			&
			n_{L}\,F^{RL} + \kappa_{RL}\,G^{LL}_{22}
			&
			n_{L}\,G^{RR}_{11} + \kappa_{RL}\,\bar F^{RR}
			&
			n_{L}\,F^{RR} + \kappa_{RL}\,G^{RR}_{22}
			\\[0.8ex]
			-\,n_{L}\,\bar F^{RL} + \kappa_{LR}^{*}\,G^{LL}_{11}
			&
			-\,n_{L}\,G^{RL}_{22} + \kappa_{LR}^{*}\,F^{LL}
			&
			-\,n_{L}\,\bar F^{RR} + \kappa_{LR}^{*}\,G^{LR}_{11}
			&
			-\,n_{L}\,G^{RR}_{22} + \kappa_{LR}^{*}\,F^{LR}
		\end{pmatrix}.
	\end{equation*}
\end{subequations}
\end{widetext}
Then 
\begin{equation}
	G^r(\omega)
	=
	\left[
	\begin{pmatrix}
		\tilde\omega - \tilde{\varepsilon}_L & \tilde{\Delta}_L & 0 & \tilde{\Delta}_{C}^{(LR)} \\
		\tilde{\Delta}_L^{*} & \tilde\omega + \tilde{\varepsilon}_L & \tilde{\Delta}_{C}^{(RL)} & 0 \\
		0 & \big(\tilde{\Delta}_{C}^{(RL)}\big)^{*} & \tilde\omega - \tilde{\varepsilon}_R & \tilde{\Delta}_R \\
		\big(\tilde{\Delta}_{C}^{(LR)}\big)^{*} & 0 & \tilde{\Delta}_R^{*} & \tilde\omega + \tilde{\varepsilon}_R
	\end{pmatrix}
	\right]^{-1}.
	\label{eq:G_inv_WBL}
\end{equation}
where $\tilde\omega = \omega + i\Gamma_N/2$, 
$	\tilde{\varepsilon}_{\alpha}
	= \varepsilon_{\alpha}
	+ U_{\alpha}\,\frac{n_{\alpha}}{2}
	+ U_{LR}\,n_{\bar{\alpha}}$,
$	\tilde{\Delta}_{\alpha}
	= \frac{\Gamma_S}{2}
	+ U_{\alpha}\,\kappa_{\alpha}$, and 
$	\tilde{\Delta}_{C}^{(\alpha)}
	= \frac{\Gamma_C}{2}
	+ U_{LR}\,\kappa_{\alpha\bar{\alpha}}$.

\subsection{Comparison with Real-Time Diagrammatics at small interaction and small voltage}

In this subsection we compare the results obtained from the real-time diagrammatic approach with those derived within the Hartree--Fock Green’s-function formalism in the regime of weak interactions and small applied bias. This comparison is restricted to parameter ranges where the HF approximation is considered to be accurate, namely $U \lesssim \Gamma_N$ and $|\mu_N| \ll k_B T$. 

In Fig.~\ref{figure11} we compare the Green’s-function–based results with those obtained from the RTD  approach in the regime of small interaction strength and small voltage. In the noninteracting limit $U=0$, the HF, RTD, and exact nonequilibrium Green’s-function calculations are found to coincide for all quantities shown, including the current, zero-frequency noise, and both the classical and quantum TURs, providing a nontrivial benchmark for the different formalisms. Upon increasing the Coulomb interaction, the main effect is a shift of the transport features toward lower gate energies, while the current and noise remain quantitatively very similar within HF and RTD, even at finite $U$. In contrast, the TURs exhibit a much stronger sensitivity to interactions: their negative contributions, which signal violations of the corresponding bounds, differ substantially between the two approaches. This behavior becomes particularly apparent at larger interaction strengths, e.g., for $U/\Gamma_N \simeq 1$, where the quantum TUR obtained within HF still displays a clear violation, whereas the RTD approach predicts its complete suppression.

A similar comparison for the Cooper-pair splitter is shown in Fig.~\ref{figure12}. In the noninteracting limit $U=0$, the Hartree--Fock, real-time diagrammatic, and exact nonequilibrium Green’s-function results again coincide for all observables, including the current, noise, and both TURs. As the interaction is increased to $U/\Gamma_N = 0.3$, the current and noise remain nearly indistinguishable between HF and RTD, while noticeable differences already emerge in the violation regimes of the classical and quantum TURs. For larger interaction strength, $U/\Gamma_N \simeq 1$, small discrepancies become visible also at the level of the current and noise. In this regime, the classical TUR is violated within both approaches, although the violation is significantly stronger in the HF results, whereas the quantum TUR exhibits a violation only within the HF approximation and is fully restored within RTD. Away from the resonance regions, where the TURs attain their minima, all three approaches yield very similar results. These findings further support the conclusion that TURs constitute a particularly sensitive probe of interaction effects in superconducting hybrid systems, revealing qualitative differences between theoretical descriptions even when standard transport observables remain largely unaffected.

\bibliography{biblio}

@article{governale2008real,
  title={Real-time diagrammatic approach to transport through interacting quantum dots with normal and superconducting leads},
  author={Governale, Michele and Pala, Marco G and K{\"o}nig, J{\"u}rgen},
  journal={Phys. Rev. B},
  volume={77},
  number={13},
  pages={134513},
  year={2008},
  publisher={APS},
  doi = {10.1103/PhysRevB.77.134513}
}

@article{ohnmacht2025role,
  title={The role of charge in thermodynamic uncertainty relations},
  author={Ohnmacht, David Christian and Belzig, Wolfgang and Cuevas, Juan Carlos},
  journal={arXiv preprint arXiv:2512.20558},
  year={2025}
}

@article{konig1996zero,
  title={Zero-bias anomalies and boson-assisted tunneling through quantum dots},
  author={K{\"o}nig, J{\"u}rgen and Schoeller, Herbert and Sch{\"o}n, Gerd},
  journal={Phys. Rev. Lett.},
  volume={76},
  number={10},
  pages={1715},
  year={1996},
  publisher={APS},
  doi = {10.1103/PhysRevLett.76.1715}
}

@article{flindt2010counting,
  title={{Counting statistics of transport through Coulomb blockade nanostructures: High-order cumulants and non-Markovian effects}},
  author={Flindt, Christian and Novotn{\`y}, Tom{\'a}{\v{s}} and Braggio, Alessandro and Jauho, Antti-Pekka},
  journal={Phys. Rev. B},
  volume={82},
  number={15},
  pages={155407},
  year={2010},
  publisher={APS},
  doi = {10.1103/PhysRevB.82.155407}
}

@article{droste2012josephson,
  title={Josephson current through interacting double quantum dots with spin--orbit coupling},
  author={Droste, Stephanie and Andergassen, Sabine and Splettstoesser, Janine},
  journal={Journal of Physics: Condensed Matter},
  volume={24},
  number={41},
  pages={415301},
  year={2012},
  publisher={IOP Publishing},
  doi = {10.1088/0953-8984/24/41/415301}
}

@article{mayo2025thermodynamic,
  title = {Quantum thermodynamic uncertainty relation and macroscopic superconducting coherence},
  author = {Mayo, Franco and Sobrino, Nahual and Fazio, Rosario and Taddei, Fabio and Governale, Michele},
  journal = {Phys. Rev. Res.},
  pages = {--},
  year = {2026},
  month = {Jan},
  publisher = {American Physical Society},
  doi = {10.1103/tfsv-fjzd},
  url = {https://link.aps.org/doi/10.1103/tfsv-fjzd}
}

@article{pala2007nonequilibrium,
  title={{Nonequilibrium Josephson and Andreev current through interacting quantum dots}},
  author={Pala, Marco G and Governale, Michele and K{\"o}nig, J{\"u}rgen},
  journal={New Journal of Physics},
  volume={9},
  number={8},
  pages={278},
  year={2007},
  publisher={IOP Publishing},
  doi = {10.1088/1367-2630/9/8/278}
}

@article{barato2015thermodynamic,
  title={Thermodynamic uncertainty relation for biomolecular processes},
  author={Barato, Andre C and Seifert, Udo},
  journal={Phys. Rev. Lett.},
  volume={114},
  number={15},
  pages={158101},
  year={2015},
  publisher={APS},
  doi = {10.1103/PhysRevLett.114.158101}
}

@article{gingrich2016dissipation,
  title={Dissipation bounds all steady-state current fluctuations},
  author={Gingrich, Todd R and Horowitz, Jordan M and Perunov, Nikolay and England, Jeremy L},
  journal={Phys. Rev. Lett.},
  volume={116},
  number={12},
  pages={120601},
  year={2016},
  publisher={APS},
  doi = {10.1103/PhysRevLett.116.120601},
}

@article{horowitz2020thermodynamic,
  title={Thermodynamic uncertainty relations constrain non-equilibrium fluctuations},
  author={Horowitz, Jordan M and Gingrich, Todd R},
  journal={Nature Physics},
  volume={16},
  number={1},
  pages={15--20},
  year={2020},
  publisher={Nature Publishing Group UK London},
doi = {10.1038/s41567-019-0702-6},
}

@article{Pietzonka2018,
  title = {Universal Trade-Off between Power, Efficiency, and Constancy in Steady-State Heat Engines},
  author = {Pietzonka, Patrick and Seifert, Udo},
  journal = {Phys. Rev. Lett.},
  volume = {120},
  issue = {19},
  pages = {190602},
  numpages = {6},
  year = {2018},
  month = {May},
  publisher = {American Physical Society},
  doi = {10.1103/PhysRevLett.120.190602},
  url = {https://link.aps.org/doi/10.1103/PhysRevLett.120.190602}
}

@article{sobrino2021thermoelectric,
  title={Thermoelectric transport within density functional theory},
  author={Sobrino, N and Eich, F and Stefanucci, G and D'Agosta, R and Kurth, S},
  journal={Phys. Rev. B},
  volume={104},
  number={12},
  year={2021},
  publisher={American Physical Society},
doi = {10.1103/PhysRevB.104.125115}
}

@article{Guarnieri2019,
  title = {Thermodynamics of precision in quantum nonequilibrium steady states},
  author = {Guarnieri, Giacomo and Landi, Gabriel T. and Clark, Stephen R. and Goold, John},
  journal = {Phys. Rev. Res.},
  volume = {1},
  issue = {3},
  pages = {033021},
  numpages = {13},
  year = {2019},
  month = {Oct},
  publisher = {American Physical Society},
  doi = {10.1103/PhysRevResearch.1.033021},
  url = {https://link.aps.org/doi/10.1103/PhysRevResearch.1.033021}
}

@Article{Kheradsoud2019,
AUTHOR = {Kheradsoud, Sara and Dashti, Nastaran and Misiorny, Maciej and Potts, Patrick P. and Splettstoesser, Janine and Samuelsson, Peter},
TITLE = {Power, Efficiency and Fluctuations in a Quantum Point Contact as Steady-State Thermoelectric Heat Engine},
JOURNAL = {Entropy},
VOLUME = {21},
YEAR = {2019},
NUMBER = {8},
ARTICLE-NUMBER = {777},
URL = {https://www.mdpi.com/1099-4300/21/8/777},
PubMedID = {33267490},
ISSN = {1099-4300},
DOI = {10.3390/e21080777}
}

@article{Ptaszynski2018,
  title = {Coherence-enhanced constancy of a quantum thermoelectric generator},
  author = {Ptaszy\ifmmode \acute{n}\else \'{n}\fi{}ski, Krzysztof},
  journal = {Phys. Rev. B},
  volume = {98},
  issue = {8},
  pages = {085425},
  numpages = {11},
  year = {2018},
  month = {Aug},
  publisher = {American Physical Society},
  doi = {10.1103/PhysRevB.98.085425},
  url = {https://link.aps.org/doi/10.1103/PhysRevB.98.085425}
}

@article{Ehrlich2021,
  title = {Broadband frequency filters with quantum dot chains},
  author = {Ehrlich, Tilmann and Schaller, Gernot},
  journal = {Phys. Rev. B},
  volume = {104},
  issue = {4},
  pages = {045424},
  numpages = {12},
  year = {2021},
  month = {Jul},
  publisher = {American Physical Society},
  doi = {10.1103/PhysRevB.104.045424},
  url = {https://link.aps.org/doi/10.1103/PhysRevB.104.045424}
}

@article{Brandner2018,
  title = {Thermodynamic Bounds on Precision in Ballistic Multiterminal Transport},
  author = {Brandner, Kay and Hanazato, Taro and Saito, Keiji},
  journal = {Phys. Rev. Lett.},
  volume = {120},
  issue = {9},
  pages = {090601},
  numpages = {6},
  year = {2018},
  month = {Mar},
  publisher = {American Physical Society},
  doi = {10.1103/PhysRevLett.120.090601},
  url = {https://link.aps.org/doi/10.1103/PhysRevLett.120.090601}
}

@article{Prech2023,
  title = {Entanglement and thermokinetic uncertainty relations in coherent mesoscopic transport},
  author = {Prech, Kacper and Johansson, Philip and Nyholm, Elias and Landi, Gabriel T. and Verdozzi, Claudio and Samuelsson, Peter and Potts, Patrick P.},
  journal = {Phys. Rev. Res.},
  volume = {5},
  issue = {2},
  pages = {023155},
  numpages = {22},
  year = {2023},
  month = {Jun},
  publisher = {American Physical Society},
  doi = {10.1103/PhysRevResearch.5.023155},
  url = {https://link.aps.org/doi/10.1103/PhysRevResearch.5.023155}
}

@article{eldridge2010superconducting,
  title={Superconducting proximity effect in interacting double-dot systems},
  author={Eldridge, James and Pala, Marco G and Governale, Michele and K{\"o}nig, J{\"u}rgen},
  journal={Phys. Rev. B},
  volume={82},
  number={18},
  pages={184507},
  year={2010},
  publisher={APS},
doi = {10.1103/PhysRevB.82.184507},
}

@article{taddei2023thermodynamic,
  title={Thermodynamic uncertainty relations for systems with broken time reversal symmetry: The case of superconducting hybrid systems},
  author={Taddei, Fabio and Fazio, Rosario},
  journal={Phys. Rev. B},
  volume={108},
  number={11},
  pages={115422},
  year={2023},
  publisher={APS},
doi={10.1103/PhysRevB.108.115422},
}

@article{braggio2011superconducting,
  title={Superconducting proximity effect in interacting quantum dots revealed by shot noise},
  author={Braggio, Alessandro and Governale, Michele and Pala, Marco G and K{\"o}nig, J{\"u}rgen},
  journal={Solid State Communications},
  volume={151},
  number={2},
  pages={155--158},
  year={2011},
  publisher={Elsevier},
doi = {10.1016/j.ssc.2010.10.043},
}

@article{Benenti2017,
title = {Fundamental aspects of steady-state conversion of heat to work at the nanoscale},
journal = {Physics Reports},
volume = {694},
pages = {1-124},
year = {2017},
issn = {0370-1573},
doi = {https://doi.org/10.1016/j.physrep.2017.05.008},
url = {https://www.sciencedirect.com/science/article/pii/S0370157317301540},
author = {Giuliano Benenti and Giulio Casati and Keiji Saito and Robert S. Whitney}
}

@article{Saryal2019,
  title={Thermodynamic uncertainty relation in thermal transport},
  author={Saryal, Sushant and Friedman, Hava Meira and Segal, Dvira and Agarwalla, Bijay Kumar},
  journal={Phys. Rev. E},
  volume={100},
  number={4},
  pages={042101},
  year={2019},
  publisher={APS},
  doi = {10.1103/PhysRevE.100.042101},
}

@article{Saryal2021,
  title = {Universal Bounds on Fluctuations in Continuous Thermal Machines},
  author = {Saryal, Sushant and Gerry, Matthew and Khait, Ilia and Segal, Dvira and Agarwalla, Bijay Kumar},
  journal = {Phys. Rev. Lett.},
  volume = {127},
  issue = {19},
  pages = {190603},
  numpages = {7},
  year = {2021},
  month = {Nov},
  publisher = {American Physical Society},
  doi = {10.1103/PhysRevLett.127.190603},
  url = {https://link.aps.org/doi/10.1103/PhysRevLett.127.190603}
}

@article{Saryal2022,
  title = {Bounds on fluctuations for machines with broken time-reversal symmetry: A linear response study},
  author = {Saryal, Sushant and Mohanta, Sandipan and Agarwalla, Bijay Kumar},
  journal = {Phys. Rev. E},
  volume = {105},
  issue = {2},
  pages = {024129},
  numpages = {9},
  year = {2022},
  month = {Feb},
  publisher = {American Physical Society},
  doi = {10.1103/PhysRevE.105.024129},
  url = {https://link.aps.org/doi/10.1103/PhysRevE.105.024129}
}

@article{Kamijima2021,
  title = {Higher-order efficiency bound and its application to nonlinear nanothermoelectrics},
  author = {Kamijima, Takuya and Otsubo, Shun and Ashida, Yuto and Sagawa, Takahiro},
  journal = {Phys. Rev. E},
  volume = {104},
  issue = {4},
  pages = {044115},
  numpages = {8},
  year = {2021},
  month = {Oct},
  publisher = {American Physical Society},
  doi = {10.1103/PhysRevE.104.044115},
  url = {https://link.aps.org/doi/10.1103/PhysRevE.104.044115}
}

@article{Palmqvist2024,
  title={Kinetic uncertainty relations for quantum transport},
  author={Palmqvist, Didrik and Tesser, Ludovico and Splettstoesser, Janine},
  journal={Phys. Rev. Lett.},
  volume={135},
  number={16},
  pages={166302},
  year={2025},
  publisher={APS},
doi = {10.1103/kvdn-skn1},
}

@misc{Zhang2025,
      title={Thermodynamic uncertainty relations for three-terminal systems with broken time-reversal symmetry}, 
      author={Yanchao Zhang and Shanhe Su},
      year={2025},
      eprint={2503.13851},
      archivePrefix={arXiv},
      primaryClass={cond-mat.stat-mech},
      url={https://arxiv.org/abs/2503.13851}, 
}

@article{Brandner2025,
  title = {Thermodynamic Uncertainty Relations for Coherent Transport},
  author = {Brandner, Kay and Saito, Keiji},
  journal = {Phys. Rev. Lett.},
  volume = {135},
  issue = {4},
  pages = {046302},
  numpages = {7},
  year = {2025},
  month = {Jul},
  publisher = {American Physical Society},
  doi = {10.1103/6nww-8wcp},
  url = {https://link.aps.org/doi/10.1103/6nww-8wcp}
}

@article{Misaki2021,
  title = {{Theory of the nonreciprocal Josephson effect}},
  author = {Misaki, Kou and Nagaosa, Naoto},
  journal = {Phys. Rev. B},
  volume = {103},
  issue = {24},
  pages = {245302},
  numpages = {10},
  year = {2021},
  month = {Jun},
  publisher = {American Physical Society},
  doi = {10.1103/PhysRevB.103.245302},
  url = {https://link.aps.org/doi/10.1103/PhysRevB.103.245302}
}

@article{Manzano2023Oct,
  author = {Manzano, Gonzalo and L{\ifmmode\acute{o}\else\'{o}\fi}pez, Rosa},
  title = {{Quantum-enhanced performance in superconducting Andreev reflection engines}},
  journal = {Phys. Rev. Res.},
  volume = {5},
  number = {4},
  pages = {043041},
  year = {2023},
  month = oct,
  publisher = {American Physical Society},
  doi = {10.1103/PhysRevResearch.5.043041}
}

@article{Lopez2023,
  title = {Optimal superconducting hybrid machine},
  author = {L\'opez, Rosa and Lim, Jong Soo and Kim, Kun Woo},
  journal = {Phys. Rev. Res.},
  volume = {5},
  issue = {1},
  pages = {013038},
  numpages = {10},
  year = {2023},
  month = {Jan},
  publisher = {American Physical Society},
  doi = {10.1103/PhysRevResearch.5.013038},
  url = {https://link.aps.org/doi/10.1103/PhysRevResearch.5.013038}
}

@article{Ohnmacht2024,
  title = {Thermodynamic uncertainty relations in superconducting junctions},
  author = {Ohnmacht, David Christian and Cuevas, Juan Carlos and Belzig, Wolfgang and L\'opez, Rosa and Lim, Jong Soo and Kim, Kun Woo},
  journal = {Phys. Rev. Res.},
  volume = {7},
  issue = {1},
  pages = {L012075},
  numpages = {6},
  year = {2025},
  month = {Mar},
  publisher = {American Physical Society},
  doi = {10.1103/PhysRevResearch.7.L012075},
  url = {https://link.aps.org/doi/10.1103/PhysRevResearch.7.L012075}
}

@article{Proesmans2019,
doi = {10.1088/1742-5468/ab14da},
url = {https://dx.doi.org/10.1088/1742-5468/ab14da},
year = {2019},
month = {may},
publisher = {IOP Publishing and SISSA},
volume = {2019},
number = {5},
pages = {054005},
author = {Proesmans, Karel and Horowitz, Jordan M},
title = {Hysteretic thermodynamic uncertainty relation for systems with broken time-reversal symmetry},
journal = {Journal of Statistical Mechanics: Theory and Experiment}
}

@article{Wozny2025,
  title = {Current noise in quantum dot thermoelectric engines},
  author = {Wozny, Simon and Leijnse, Martin},
  journal = {Phys. Rev. B},
  volume = {111},
  issue = {7},
  pages = {075422},
  numpages = {10},
  year = {2025},
  month = {Feb},
  publisher = {American Physical Society},
  doi = {10.1103/PhysRevB.111.075422},
  url = {https://link.aps.org/doi/10.1103/PhysRevB.111.075422}
}

@article{Potanina2021,
  title = {Thermodynamic bounds on coherent transport in periodically driven conductors},
  author = {Potanina, Elina and Flindt, Christian and Moskalets, Michael and Brandner, Kay},
  journal = {Phys. Rev. X},
  volume = {11},
  issue = {2},
  pages = {021013},
  numpages = {26},
  year = {2021},
  month = {Apr},
  publisher = {American Physical Society},
  doi = {10.1103/PhysRevX.11.021013},
  url = {https://link.aps.org/doi/10.1103/PhysRevX.11.021013}
}

@article{Lu2022,
  title = {Geometric thermodynamic uncertainty relation in a periodically driven thermoelectric heat engine},
  author = {Lu, Jincheng and Wang, Zi and Peng, Jiebin and Wang, Chen and Jiang, Jian-Hua and Ren, Jie},
  journal = {Phys. Rev. B},
  volume = {105},
  issue = {11},
  pages = {115428},
  numpages = {10},
  year = {2022},
  month = {Mar},
  publisher = {American Physical Society},
  doi = {10.1103/PhysRevB.105.115428},
  url = {https://link.aps.org/doi/10.1103/PhysRevB.105.115428}
}

@article{anantram1996current,
  author  = {M. P. Anantram and S. Datta},
  title   = {{Current fluctuations in mesoscopic systems with Andreev scattering}},
  journal = {Phys. Rev. B},
  volume  = {53},
  number  = {24},
  pages   = {16390--16402},
  year    = {1996},
  doi     = {10.1103/PhysRevB.53.16390}
}

@article{Kaneko2016,
doi = {10.1088/0957-0233/27/3/032001},
url = {https://doi.org/10.1088/0957-0233/27/3/032001},
year = {2016},
month = {feb},
publisher = {IOP Publishing},
volume = {27},
number = {3},
pages = {032001},
author = {Kaneko, Nobu-Hisa and Nakamura, Shuji and Okazaki, Yuma},
title = {A review of the quantum current standard},
journal = {Meas. Sci. Technol.}
}

@article{Kaneko2024,
doi = {10.1088/1361-6501/ad03a2},
url = {https://doi.org/10.1088/1361-6501/ad03a2},
year = {2023},
month = {oct},
publisher = {IOP Publishing},
volume = {35},
number = {1},
pages = {011001},
author = {Kaneko, Nobu-Hisa and Tanaka, Takahiro and Okazaki, Yuma},
title = {Perspectives of the generation and measurement of small electric currents},
journal = {Meas. Sci. Technol.}
}

@article{Bolech2007,
  title = {{Observing Majorana bound states in $p$-wave superconductors using noise measurements in tunneling experiments}},
  author = {Bolech, C. J. and Demler, Eugene},
  journal = {Phys. Rev. Lett.},
  volume = {98},
  issue = {23},
  pages = {237002},
  numpages = {4},
  year = {2007},
  month = {Jun},
  publisher = {American Physical Society},
  doi = {10.1103/PhysRevLett.98.237002},
  url = {https://link.aps.org/doi/10.1103/PhysRevLett.98.237002}
}

@article{Arrachea2024,
  title = {{Signatures of triplet superconductivity in $\ensuremath{\nu}=2$ chiral Andreev states}},
  author = {Arrachea, Liliana and Yeyati, Alfredo Levy and Balseiro, C. A.},
  journal = {Phys. Rev. B},
  volume = {109},
  issue = {6},
  pages = {064519},
  numpages = {9},
  year = {2024},
  month = {Feb},
  publisher = {American Physical Society},
  doi = {10.1103/PhysRevB.109.064519},
  url = {https://link.aps.org/doi/10.1103/PhysRevB.109.064519}
}

@article{Cangemi2024,
title = {Quantum engines and refrigerators},
journal = {Physics Reports},
volume = {1087},
pages = {1-71},
year = {2024},
note = {Quantum engines and refrigerators},
issn = {0370-1573},
doi = {https://doi.org/10.1016/j.physrep.2024.07.001},
url = {https://www.sciencedirect.com/science/article/pii/S0370157324002710},
author = {Loris Maria Cangemi and Chitrak Bhadra and Amikam Levy}
}

@article{Balduque2025,
  title={Quantum thermocouples: Nonlocal conversion and control of heat in nanostructures},
  author={Balduque, Jos{\'e} and S{\'a}nchez, Rafael},
  journal = {Eur. Phys. J. Spec. Top.},
  pages={1--31},
  year={2026},
  publisher={Springer},
doi = {10.1140/epjs/s11734-025-02107-8},
}

@article{sobrino2025thermoelectric,
  title={Thermoelectric properties of interacting double quantum dots},
  author={Sobrino, Nahual},
  journal={Phys. Rev. B},
  volume={112},
  number={23},
  pages={235101},
  year={2025},
  publisher={APS},
 doi = {10.1103/c8pl-h1jh},
}

@article{Muzykantskii1994,
  title = {Quantum shot noise in a normal-metal--superconductor point contact},
  author = {Muzykantskii, B. A. and Khmelnitskii, D. E.},
  journal = {Phys. Rev. B},
  volume = {50},
  issue = {6},
  pages = {3982--3987},
  numpages = {0},
  year = {1994},
  month = {Aug},
  publisher = {American Physical Society},
  doi = {10.1103/PhysRevB.50.3982},
  url = {https://link.aps.org/doi/10.1103/PhysRevB.50.3982}
}

@article{Martin1996,
title = {{Wave packet approach to noise in N--S junctions}},
journal = {Physics Letters A},
volume = {220},
number = {1},
pages = {137-142},
year = {1996},
issn = {0375-9601},
doi = {https://doi.org/10.1016/0375-9601(96)00484-7},
url = {https://www.sciencedirect.com/science/article/pii/0375960196004847},
author = {Thierry Martin}
}

@article{deJong1994,
  title = {Doubled shot noise in disordered normal-metal--superconductor junctions},
  author = {de Jong, M. J. M. and Beenakker, C. W. J.},
  journal = {Phys. Rev. B},
  volume = {49},
  issue = {22},
  pages = {16070--16073},
  numpages = {0},
  year = {1994},
  month = {Jun},
  publisher = {American Physical Society},
  doi = {10.1103/PhysRevB.49.16070},
  url = {https://link.aps.org/doi/10.1103/PhysRevB.49.16070}
}

@article{Khlus1987,
  title={Current and voltage fluctuations in microjunctions between normal metals and superconductors},
  author={Khlus, VA},
  journal={Zh. Eksp. Teor. Fiz},
  volume={93},
  pages={2179},
  year={1987},
}

@article{Golub2011,
  title = {{Shot noise in a Majorana fermion chain}},
  author = {Golub, Anatoly and Horovitz, Baruch},
  journal = {Phys. Rev. B},
  volume = {83},
  issue = {15},
  pages = {153415},
  numpages = {4},
  year = {2011},
  month = {Apr},
  publisher = {American Physical Society},
  doi = {10.1103/PhysRevB.83.153415},
  url = {https://link.aps.org/doi/10.1103/PhysRevB.83.153415}
}

@article{WU2012,
  title = {{Tunneling transport through superconducting wires with Majorana bound states}},
  author = {Wu, B. H. and Cao, J. C.},
  journal = {Phys. Rev. B},
  volume = {85},
  issue = {8},
  pages = {085415},
  numpages = {7},
  year = {2012},
  month = {Feb},
  publisher = {American Physical Society},
  doi = {10.1103/PhysRevB.85.085415},
  url = {https://link.aps.org/doi/10.1103/PhysRevB.85.085415}
}

@article{Lu2012,
  title = {{Nonlocal noise cross correlation mediated by entangled Majorana fermions}},
  author = {L\"u, Hai-Feng and Lu, Hai-Zhou and Shen, Shun-Qing},
  journal = {Phys. Rev. B},
  volume = {86},
  issue = {7},
  pages = {075318},
  numpages = {7},
  year = {2012},
  month = {Aug},
  publisher = {American Physical Society},
  doi = {10.1103/PhysRevB.86.075318},
  url = {https://link.aps.org/doi/10.1103/PhysRevB.86.075318}
}

@article{Weiss2017,
  title = {Odd-triplet superconductivity in single-level quantum dots},
  author = {Weiss, Stephan and K\"onig, J\"urgen},
  journal = {Phys. Rev. B},
  volume = {96},
  issue = {6},
  pages = {064529},
  numpages = {11},
  year = {2017},
  month = {Aug},
  publisher = {American Physical Society},
  doi = {10.1103/PhysRevB.96.064529},
  url = {https://link.aps.org/doi/10.1103/PhysRevB.96.064529}
}

@article{Seoane2020,
  title = {{Signatures of odd-frequency pairing in the Josephson junction current noise}},
  author = {Seoane Souto, Rub\'en and Kuzmanovski, Dushko and Balatsky, Alexander V.},
  journal = {Phys. Rev. Res.},
  volume = {2},
  issue = {4},
  pages = {043193},
  numpages = {11},
  year = {2020},
  month = {Nov},
  publisher = {American Physical Society},
  doi = {10.1103/PhysRevResearch.2.043193},
  url = {https://link.aps.org/doi/10.1103/PhysRevResearch.2.043193}
}

@Article{Gonzalez2025,
author="C. González-Ruano and C. Shen and P. Tuero and C. Tiusan and Y. Lu and J. E. Han and I. \v{Z}uti\'c and F. G. Aliev",
title="Giant shot noise in superconductor/ferromagnet junctions with orbital-symmetry-controlled spin-orbit coupling",
journal="Nat. Commun.",
year="2025",
volume="16",
pages="9524",
doi="10.1038/s41467-025-64493-w",
}

@book{Degennesbook,
  title        = {Superconductivity of Metals and Alloys},
  author       = {de Gennes, Pierre-Gilles},
  year         = {1966},
  publisher    = {W. A. Benjamin},
  address      = {New York},
  note         = {Reprinted by CRC Press (1999, 2018)}
}

@article{Blonder1982,
  title        = {Transition from metallic to tunneling regimes in superconducting microconstrictions: Excess current, charge imbalance, and supercurrent},
  author       = {Blonder, G.~E. and Tinkham, M. and Klapwijk, T.~M.},
  journal      = {Phys. Rev. B},
  volume       = {25},
  number       = {7},
  pages        = {4515--4532},
  year         = {1982},
  publisher    = {American Physical Society},
  doi          = {10.1103/PhysRevB.25.4515}
}

@article{Fazio1998,
  title = {{Resonant Andreev Tunneling in Strongly Interacting Quantum Dots}},
  author = {Fazio, Rosario and Raimondi, Roberto},
  journal = {Phys. Rev. Lett.},
  volume = {80},
  issue = {13},
  pages = {2913--2916},
  numpages = {0},
  year = {1998},
  month = {Mar},
  publisher = {American Physical Society},
  doi = {10.1103/PhysRevLett.80.2913},
  url = {https://link.aps.org/doi/10.1103/PhysRevLett.80.2913}
}

@article{Clerk2000,
  title = {{Andreev scattering and the Kondo effect}},
  author = {Clerk, Aashish A. and Ambegaokar, Vinay and Hershfield, Selman},
  journal = {Phys. Rev. B},
  volume = {61},
  issue = {5},
  pages = {3555--3562},
  numpages = {0},
  year = {2000},
  month = {Feb},
  publisher = {American Physical Society},
  doi = {10.1103/PhysRevB.61.3555},
  url = {https://link.aps.org/doi/10.1103/PhysRevB.61.3555}
}
\end{document}